\documentclass[review]{elsarticle}

\usepackage{lineno,hyperref}

\usepackage{color}

\usepackage{amsmath}

\usepackage{subfig}
\usepackage{float}
\usepackage{array, threeparttable}
\usepackage{amssymb}
\usepackage{overpic}               
\modulolinenumbers[5]

\journal{Journal of \LaTeX\ Templates}

\begin{document}
	
	
	\begin{frontmatter}
		
		\title{Hypersonic wave wall flow based on gas kinetic method}

		\author[mymainaddress]{Yining Yang}
		\author[mymainaddress]{Rui Zhang}
		\author[mymainaddress]{Jianfeng Chen}
            \author[mymainaddress]{Sha Liu}
            \author[mymainaddress]{Congshan Zhuo}
		\author[mysecondaryaddress]{Weibo Hu}
            \author[mymainaddress]{Chengwen Zhong\corref{mycorrespondingauthor}}
		\cortext[mycorrespondingauthor]{Corresponding author}
		\ead{zhongcw@nwpu.edu.cn.}
		
		\address[mymainaddress]{School of Aeronautics, Northwestern Polytechnical University, Xi'an, Shaanxi 710072, China}
		\address[mysecondaryaddress]{State Key Laboratory of Aerodynamics, Mianyang, 621000, China}
		
		\begin{abstract}
			The transition of hypersonic boundary layer can lead to a several-fold increase in surface heat flux and skin friction for the aircraft, significantly impacting its flight performance. The corrugated wall, as a passive control method for boundary layer flow, also serves as a type of wall microstructure, making its study on the local rarefaction effect of considerable engineering significance. In this study, we employed the conservative discrete unified gas dynamic scheme and utilized a domain-wide numerical simulation method. Initially, we simulated the hypersonic flat plate flow with different depths of corrugated walls under the conditions of incoming flow Mach number of 6 and Reynolds number of ${{10}^{7}}$. Subsequently, we investigated the effects of corrugated walls, including flat plate corrugated walls and wedge corrugated walls, under varying Reynolds numbers for an incoming flow Mach number of 6, and discussed the impact of local rarefaction effect of corrugated walls under different Reynolds numbers. By using the local Knudsen number as the criterion, we found that under these conditions, the occurrence of local rarefaction effect near the corrugated wall due to consecutive failures does not take place when the incoming Reynolds number reaches ${{10}^{7}}$ or ${{10}^{6}}$. However, when the incoming Reynolds number drops to ${{10}^{5}}$, the local rarefaction effect near the corrugated wall becomes evident, with the appearance of non-equilibrium effects in translational and rotational temperatures of molecules. This phenomenon becomes more pronounced as the Reynolds number decreases further.
		\end{abstract}
		
		
	\end{frontmatter}
	
	
	\section{Introduction}
	Modeling the boundary layer around the surface of near-space vehicles is crucial for predicting the aerodynamic characteristics of the vehicles. The local flow field within the boundary layer can be extremely complex, encompassing both continuum flow and rarefied non-equilibrium flow. On one hand, hypersonic vehicles, upon entering the atmosphere, traverse different flow regimes due to variations in atmospheric density with altitude, including continuum, slip, transitional, and free-molecular flow regimes. On the other hand, the flow around the nose of a hypersonic vehicle is characterized by strong compression, resulting in continuum flow, while the flow near the vehicle's leeward region experiences expansion, leading to reduced flow density and rarefied flow. Local non-equilibrium effects may arise due to the localized rarefaction of gases or extreme spatiotemporal variations in velocity. Therefore, achieving a numerically efficient and accurate simulation of such supersonic flows poses significant challenges for practical applications.\par
	For the classical supersonic flat plate boundary layer problem, the interaction of shock waves with the boundary layer causes the breakdown of the classical laminar boundary layer theory as the flow approaches upstream from downstream of the flat plate~\cite{doi:10.2514/1.J053168}. Further upstream, the Navier-Stokes equations based on the continuity assumption also become invalid, and slip flow, transitional flow, and free-molecular flow gradually emerge. For these more complex flow scenarios, there is a lack of available theories for quantitative prediction, leading to an overreliance on numerical methods and empirically fitted formulas in practical applications. \par
	Since the 1980s, there has been an increasing amount of CFD research on this problem. For instance, Maslov et al.~\cite{maslov_viscous_1999} proposed the full viscous shock layer equations and compared the computed density distribution and shock angles with experimental data, showing good agreement at high and moderate Reynolds numbers. Gokcen et al.~\cite{ref3} introduced the slip boundary condition to the NS equations for transitional flows, simplifying it to a known slip condition for low Knudsen numbers, enabling correct shear stress and heat transfer predictions in transitional flow regions for flat plate simulations. Bird~\cite{ref4} used DSMC to study shock-boundary layer interactions in low-density supersonic flows, simulating representative flat plate, wedge, and cone boundary layers and determining the maximum pressure, shear stress, and heat transfer in the shock formation regions for each case. Recently, Greenshields et al.~\cite{GREENSHIELDS201280} and Lofthouse et al.~\cite{doi:10.2514/1.31280} applied improved non-continuum boundary conditions in simulations that showed good agreement with DSMC results for high hypersonic rarefied gas flows. Tsuboi and Matsumoto~\cite{doi:10.2514/1.10950} studied the near-equilibrium state near the leading edge in a three-dimensional model through experiments and DSMC simulations, validating the significant imbalance between translational and rotational temperatures near the leading edge of a flat plate. In addition to particle methods and slip conditions, another approach to modeling rarefied gas effects is through the use of nonlinear constitutive relations, such as the numerical solution of the Burnett equations by Tannehill~\cite{10.1063/1.861304} and Titarev~\cite{titarev_numerical_2005}, which cover continuum, slip flow, and partial transitional flows with Knudsen numbers less than 1.\par
	Microscale structures and surface roughness on the aircraft surface significantly affect the flow characteristics near the surface and the skin friction of the aircraft, particularly under rarefied conditions where the wall effects become more prominent~\cite{doi:10.1080/00411457108231440}. Several scientific issues exist at different levels in this context:
	Nonlinear transport characteristics of the near-wall region's Knudsen layer and appropriate slip boundary models.
	Requirements of gas-surface interaction (GSI) models~\cite{verbeek_smoluchowski_2018} considering surface geometric multiscale features and their impact on wall accommodation coefficients.
	Influence of wall microstructures on near-wall flow and quantitative effects on wall friction.
	Impact of surface microscale structures on the magnitude of near-wall heat flux and their influence on the overall aerodynamic heating distribution and thermal protection design of the aircraft.
	Coupling effects between multiscale flows and multiscale structures on the wall, addressing the influencing mechanisms on the flow and the mechanisms governing key aerodynamic parameters for typical hypersonic configurations.
	These scientific inquiries are critical for understanding and addressing the intricate relationship between microscale structures, surface roughness, and flow behavior near the aircraft surface. They hold the potential to enhance our knowledge and optimize the design of aircraft surfaces, ultimately reducing skin friction and improving overall aerodynamic performance.\par
	The study aims to investigate the flow mechanism of rough wall surfaces by simplifying them into flat plates with cavities or gaps. For the problem of rarefied flow in gaps, Palharini et al.~\cite{ref12} conducted computational studies on rarefied hypersonic cavity flows using the DSMC method. The study focused on the impact of changes in the length-to-depth ratio (L/D) of the cavity on aerodynamic surface quantities. Palharini~\cite{doi:10.2514/1.A32746} further extended the study of rarefied hypersonic corner flows into three dimensions, revealing significant alterations in the flow field inside and around the cavity with the addition of a third spatial dimension. Application of two-dimensional cavity solutions greatly overestimates the aerodynamic thermodynamic load, which is several times that observed in three-dimensional studies.

	Building on Palharini's work, Jin Xuhong et al.~\cite{10.13224/j.cnki.jasp.2019.01.023} utilized the DSMC method to simulate rarefied hypersonic gap flow in the transition region, considering the effects of rarefied gas (flight altitude) and three-dimensional effects on gap flow. They explored variations in the internal flow field structure and heat flux within the gap at multiple flight altitudes. Recently, they also investigated the influence of the GSI model on the flow characteristics of concave cavities and aerodynamic surface quantities (surface pressure and heat flux), exploring the variations in flow characteristics, surface pressure, and heat flux with changes in the GSI model~\cite{124118}.

	Cao et al.~\cite{10.1063/1.1871363} employed Molecular Dynamics (MD) to study gas flow within microchannels and found that the Tangential Momentum Adaptation Coefficient (TMAC) decreases exponentially with increasing temperature. Sun and Li~\cite{doi:10.1080/00268970802452020} tracked collisions between gas molecules and surfaces and statistically determined the incident and reflected velocities of gas molecules, directly calculating the gas molecule adaptation coefficient. They further studied the effects of surface configuration, surface temperature, and physical adsorption on the coefficient. Zhang Ran et al.~\cite{10.7498/aps.67.20172706} simulated gas flow driven by shear forces in transitional nanochannels, investigating the influence of channel height and gas-solid interaction potential coefficients on TMAC. Zhang Ye et al.~\cite{10.7498/aps.68.20190987} used MD to simulate the collision process of gas molecules on solid surfaces, sampling velocities to give the incident gas macroscopic velocity characteristics. They calculated the tangential momentum adaptation coefficient, normal momentum adaptation coefficient, and energy adaptation coefficient based on the average momentum and energy of the incident and reflected gas molecules, and analyzed the variations of these coefficients with macroscopic tangential and normal velocities under smooth and rough surfaces.\par
	This paper is organized as follows. Section \ref{sec2} gives a brief review of the Computational models and numerical methods. Section \ref{sec3} presents the numerical results. Finally we end up with concluding remarks. 
	
	\section{Computational models and numerical methods}
	\label{sec2}
	\subsection{Governing equation}
	\label{sec2.1}
	UGKS method uses BGK model equation of Boltzmann equation. The general expression of the equation is as follows:
	\begin{equation}\label{equ3.1}
	\frac{\partial f}{\partial t}+{{\xi }_{x}}\frac{\partial f}{\partial x}+{{\xi }_{y}}\frac{\partial f}{\partial y}+{{\xi }_{z}}\frac{\partial f}{\partial z}=\frac{{{g}^{*}}-f}{\tau },
	\end{equation}
	where ${{g}^{*}} $ is the generalized equilibrium distribution function and $\tau $ is the generalized relaxation time, on the same order as the mean collision time of molecules.In the above equation, the modeled relaxation term replaces the complete binary collision term in the Boltzmann equation. This modeling simplifies the form of the new equation and significantly reduces the computational effort compared to the complete collision term. The generalized equilibrium distribution function depends on the parameters of macroscopic physical quantities and possesses a higher entropy.\par
	Rykov developed a Rykov model considering the internal energy of molecular rotation considering diatomic molecules, and its equilibrium distribution function is as follows:
	\begin{equation}
		{{g}^{*}}=\frac{1}{{{Z}_{r}}}{{g}_{M}}+\left( 1-\frac{1}{{{Z}_{r}}} \right){{g}_{tran}},
	\end{equation}
	Where ${{g}_{tran}}$ is the translational equilibrium state of gas, that is, the Maxwell equilibrium distribution function corresponding to ${{\lambda }_{tran}}=1/(2R{{T}_{tran}})$, and ${{Z}_{r}}$ is the translational temperature. ${{Z}_{r}}$ is the number of rotational collisions, representing the number of molecular collisions that occur when a rotational collision occurs.
	
	\subsection{Numerical method}
	\label{sec2.2}
	The UGKS method discretes time, physical space, and additional velocity space. For the evolution of the distribution function at any discrete velocity point, the finite volume method is treated by integrating the governing equation in the discrete time interval $ $ and the discrete physical space unit $ $ 
		\begin{equation}
		\int_{{{t}_{n}}}^{{{t}_{n+1}}}{\int_{{{\Omega }_{i}}}{\frac{\partial f}{\partial t}d\mathbf{x}dt+\int_{{{t}_{n}}}^{{{t}_{n+1}}}{\int_{{{\Omega }_{i}}}{\mathbf{\xi }\cdot \frac{\partial f}{\partial \mathbf{x}}d\mathbf{x}dt=}}}}\int_{{{t}_{n}}}^{{{t}_{n+1}}}{\int_{{{\Omega }_{i}}}{\frac{{{g}^{*}}-f}{\tau }d\mathbf{x}dt}}.
	\end{equation}
	In the UGKS algorithm, the distribution function recorded within the discrete physical unit is the average value within this unit, so the first term on the left side of the equation is written as
		\begin{equation}
		\int_{{{t}_{n}}}^{{{t}_{n+1}}}{\int_{{{\Omega }_{i}}}{\frac{\partial f}{\partial t}d\mathbf{x}dt=\left( {{f}^{n+1}}-{{f}^{n}} \right){{\Omega }_{i}}}}.
	\end{equation}
	Use Green's formula to rewrite the volume fraction of the second term on the left side of the equation as the area fraction
	\begin{equation}
		\int_{{{t}_{n}}}^{{{t}_{n+1}}}{\int_{{{\Omega }_{i}}}{\mathbf{\xi }\cdot \frac{\partial f}{\partial \mathbf{x}}d\mathbf{x}dt=\int_{{{t}_{n}}}^{{{t}_{n+1}}}{\oint_{\partial {{\Omega }_{i}}}{{{f}_{cf}}\mathbf{\xi }\cdot d\mathbf{S}d}t}}},
\end{equation}
	where $cf$ is cell interface subscript, ${{f}_{cf}}$ is the distribution function on the cell interface; $d\mathbf{S}$ is the differential of the unit interface $\partial {{\Omega }_{i}}$. This is the flux term. The above steps are general steps for finite volume methods.\par
	The trapezoidal integral formula is used in the time integral, and the integral of the collision term at the right end of the equation can be discretized. The information is expressed as
The trapezoidal integral formula is applied in the time integral, and the integral of the collision term at the right end of the equation can be expressed by using discrete information as
	\begin{equation}
\int_{{{t}_{n}}}^{{{t}_{n+1}}}{\int_{{{\Omega }_{i}}}{\frac{{{g}^{*}}-f}{\tau }d\mathbf{x}dt}}=\Delta t{{\Omega }_{i}}\left[ {{\left( \frac{{{g}^{*}}-f}{2\tau } \right)}^{n}}+{{\left( \frac{{{g}^{*}}-f}{2\tau } \right)}^{n+1}} \right]\
	\end{equation}
After proper simplification, the evolution equation of UGKS method is written as
\begin{equation}\label{equ3.10}
	\begin{aligned}
		& {{\left( 1+2{{\tau }^{+}} \right)}^{n+1}}f_{i}^{n+1}=f_{i}^{n}+\frac{1}{{{\Omega }_{i}}}\int_{{{t}_{n}}}^{{{t}_{n+1}}}{\oint_{\partial {{\Omega }_{i}}}{{{f}_{cf}}\left( \mathbf{\xi },t \right)\mathbf{\xi }\cdot d\mathbf{S}d}t} \\ 
		& \text{                           }+\Delta t\left[ {{\left( \frac{{{g}^{*}}-f}{2\tau } \right)}^{n}}+{{\left( \frac{{{g}^{*}}}{2\tau } \right)}^{n+1}} \right] \\ 
	\end{aligned}.
\end{equation}
So far, only the distribution function ${{f}_{cf}}\left( \mathbf{\xi },t \right)$ on the interface and the equilibrium distribution function ${{({{g}^{*}})}^{n+1}}$ of n+1 time step and relaxation time ${{(\tau )}^{n+1}}$ are not determined in the evolution equation.Since ${{g}^{*}}$ and $\tau $ are known to depend on macroscopic quantities, evolutionary equations of macroscopic quantities can be constructed to update ${{g}^{*}}$ and $\tau $. The left and right sides of the evolution equation of the distribution function are multiplied by the conserved physical quantity $\mathbf{\psi }$ and integrated over the entire velocity space. Since the moment of the collision term is 0 for the conserved quantity, we obtain
\begin{equation}\label{equ.3.11}
	{{\mathbf{W}}^{n+1}}={{\mathbf{W}}^{n}}+\frac{1}{{{\Omega }_{i}}}\int_{{{t}_{n}}}^{{{t}_{n+1}}}{\oint_{\partial {{\Omega }_{i}}}{{{f}_{cf}}\left( \mathbf{\xi },t \right)\mathbf{\psi \xi }\cdot d\mathbf{S}d}t}
\end{equation}
where
\begin{equation}\label{equ12}
\mathbf{W}=\left( \rho ,\rho \mathbf{U},\frac{1}{2}\rho {{U}^{2}}+\frac{3+K}{2}\rho RT \right) 
\end{equation}
They are density, momentum density and energy density respectively. On the other hand, for the model equation containing non-conserved macroscopic physical quantities in the equilibrium state, the corresponding non-conserved macroscopic physical quantities should also be included in the above macroscopic evolution equation. Take Shakhov model as an example, its equilibrium state contains non-conserved calorimetric flow $ \mathbf{q}$. Similar to the procedure for conserved quantities, first multiply both ends of the evolution equation by $m{{({{c}_{i}}{{c}^{2}})}^{n+1}}$ and integrate over the entire velocity space. Since $m{{({{c}_{i}}{{c}^{2}})}^{n+1}}$ is a non-conserved quantity, there is no simplification of moment 0 with the collision term, resulting in
\begin{equation}
\begin{aligned}
	& \left( 1+\frac{\Delta t\Pr }{2{{\tau }^{n+1}}} \right)\left( 2q_{i}^{n+1} \right)=\left\langle m{{\left( {{c}_{i}}{{c}^{2}} \right)}^{n+1}},{{f}^{n}} \right\rangle +\frac{1}{{{\Omega }_{i}}}\int_{{{t}_{n}}}^{{{t}_{n+1}}}{\oint_{\partial {{\Omega }_{i}}}{{{f}_{cf}}\left( \mathbf{\xi },t \right)m{{\left( {{c}_{i}}{{c}^{2}} \right)}^{n+1}}\mathbf{\xi }\cdot d\mathbf{S}d}t} \\ 
	& \text{                           }+\frac{\Delta t}{2{{\tau }^{n}}}\left\langle m{{\left( {{c}_{i}}{{c}^{2}} \right)}^{n+1}},\left( g_{Shakhov}^{n}-{{f}^{n}} \right) \right\rangle  \\ 
\end{aligned}
\end{equation}
Due to heat flow:
\begin{equation}
{{q}_{i}}=\left\langle \frac{1}{2}m{{c}_{i}}{{c}^{2}},f \right\rangle
\end{equation}
Depends on the macroscopic velocity $\mathbf{c}=\mathbf{\xi }-\mathbf{U}$, and the macroscopic velocity changes with time, so the expression of heat flow needs to be further expanded as:
\begin{equation}
{{q}_{i}}=\frac{1}{2}m\left\langle {{c}_{i}}{{c}^{2}},f \right\rangle =\frac{1}{2}m\left\langle {{c}_{i}}^{3}+{{c}_{i}}c_{j}^{2}+{{c}_{i}}c_{k}^{2},f \right\rangle
\end{equation}
Put $\mathbf{c}=\mathbf{\xi }-\mathbf{U}$ into the inner product of the above formula and arrange it in order for:
		\begin{equation}
			\begin{aligned}
				  & \left\langle {{c}_{i}}{{c}^{2}},f \right\rangle =\left\langle {{\xi }_{i}}^{3}+{{\xi }_{i}}\xi _{j}^{2}+{{\xi }_{i}}\xi _{k}^{2},f \right\rangle +\left\langle U_{i}^{3}-U_{j}^{2}{{U}_{i}}-U_{k}^{2}{{U}_{i}},f \right\rangle  \\ 
				& \text{                }+\left\langle \left( 3U_{i}^{2}+U_{j}^{2}+U_{k}^{2} \right){{\xi }_{i}}+2{{U}_{i}}{{U}_{j}}{{\xi }_{j}}+2{{U}_{i}}{{U}_{k}}{{\xi }_{k}},f \right\rangle  \\ 
				& \text{                }-\left\langle 3{{U}_{i}}{{\xi }_{i}}^{2}+2{{U}_{j}}{{\xi }_{i}}{{\xi }_{j}}+2{{U}_{k}}{{\xi }_{i}}{{\xi }_{k}}+{{U}_{i}}\xi _{j}^{2}+{{U}_{i}}\xi _{k}^{2},f \right\rangle  \\ 		
			\end{aligned}			
		\end{equation}	
On the other hand, because
		\begin{equation}
{{\mathbf{c}}^{n+1}}=\mathbf{\xi }-{{\mathbf{U}}^{n+1}}=\mathbf{\xi }-\left( {{\mathbf{U}}^{n}}+\Delta  \right)={{\mathbf{c}}^{n}}-\mathbf{\Delta }
		\end{equation}
The inner product of ${{c}_{i}}{{c}^{2}}$ of n+1 time steps and the distribution function of n time steps is expressed as
	\begin{equation}
\begin{aligned}
	& \left\langle {{\left( {{c}_{i}}{{c}^{2}} \right)}^{n+1}},{{f}^{n}} \right\rangle =\left\langle {{c}_{i}}^{3}+{{c}_{i}}c_{j}^{2}+{{c}_{i}}c_{k}^{2},{{f}^{n}} \right\rangle +\left\langle \Delta _{i}^{3}-\Delta _{j}^{2}{{\Delta }_{i}}-\Delta _{k}^{2}{{\Delta }_{i}},{{f}^{n}} \right\rangle  \\ 
	& \text{                        }+\left\langle \left( 3\Delta _{i}^{2}+\Delta _{j}^{2}+\Delta _{k}^{2} \right){{c}_{i}}+2{{\Delta }_{i}}{{\Delta }_{j}}{{c}_{j}}+2{{\Delta }_{i}}{{\Delta }_{k}}{{c}_{k}},{{f}^{n}} \right\rangle  \\ 
	& \text{                        }-\left\langle 3{{\Delta }_{i}}{{c}_{i}}^{2}+2{{\Delta }_{j}}{{c}_{i}}{{c}_{j}}+2{{\Delta }_{k}}{{c}_{i}}{{c}_{k}}+{{\Delta }_{i}}c_{j}^{2}+{{\Delta }_{i}}c_{k}^{2},{{f}^{n}} \right\rangle  \\ 
\end{aligned}
		\end{equation}
By calculating the inner product, the above formula is further written as
\begin{equation}
\begin{aligned}
	& \left\langle m{{\left( {{c}_{i}}{{c}^{2}} \right)}^{n+1}},{{f}^{n}} \right\rangle =\left( 2q_{i}^{n} \right)-\left[ {{\Delta }_{i}}\left( 3{{\rho }^{n}}R{{T}^{n}}+2{{\rho }^{n}}RT_{i}^{n} \right)-2{{\Delta }_{j}}\tau _{ij}^{n}-2{{\Delta }_{k}}\tau _{ik}^{n} \right] \\ 
	& \text{                         }+\left( \Delta _{i}^{3}-\Delta _{j}^{2}{{\Delta }_{i}}-\Delta _{k}^{2}{{\Delta }_{i}} \right){{\rho }^{n}} \\ 
\end{aligned}
\end{equation}
For the Shakhov unbalanced distribution function of step n, this inner product can be written as
\begin{equation}
\left\langle m{{\left( {{c}_{i}}{{c}^{2}} \right)}^{n+1}},g_{Shakhov}^{n} \right\rangle =\left( 1-\text{Pr} \right)\left( 2q_{i}^{n} \right)-{{\Delta }_{i}}5{{\rho }^{n}}R{{T}^{n}}+\left( \Delta _{i}^{3}-\Delta _{j}^{2}{{\Delta }_{i}}-\Delta _{k}^{2}{{\Delta }_{i}} \right){{\rho }^{n}}
\end{equation}
Where Pr is the Prandtl number, so there is
\begin{equation}
\left\langle m{{\left( {{c}_{i}}{{c}^{2}} \right)}^{n+1}},\left( g_{Shakhov}^{n}-{{f}^{n}} \right) \right\rangle =-\text{Pr}\left( 2q_{i}^{n} \right)+{{\Delta }_{i}}2{{\rho }^{n}}R\left( T_{i}^{n}-{{T}^{n}} \right)-2{{\Delta }_{j}}\tau _{ij}^{n}-2{{\Delta }_{k}}\tau _{ik}^{n}
\end{equation}
Finally, the evolution equation of heat flow can be summarized as
\begin{equation}\label{equ3.22}
\begin{aligned}
	& \left( 1+\frac{\text{Pr}\Delta t}{2{{\tau }^{n+1}}} \right)\left( 2q_{i}^{n+1} \right)=\left( 2q_{i}^{n} \right)+\left( \Delta _{i}^{3}-\Delta _{j}^{2}{{\Delta }_{i}}-\Delta _{k}^{2}{{\Delta }_{i}} \right){{\rho }^{n}} \\ 
	& \text{                          }-\left[ {{\Delta }_{i}}\left( 3{{\rho }^{n}}R{{T}^{n}}+2{{\rho }^{n}}RT_{i}^{n} \right)-2{{\Delta }_{j}}\tau _{ij}^{n}-2{{\Delta }_{k}}\tau _{ik}^{n} \right] \\ 
	& \text{                           +}\frac{1}{{{\Omega }_{i}}}\int_{{{t}_{n}}}^{{{t}_{n+1}}}{\oint_{\partial {{\Omega }_{i}}}{{{f}_{cf}}\left( \mathbf{\xi },t \right){{\left( m{{c}_{i}}{{c}^{2}} \right)}^{n+1}}\mathbf{\xi }\cdot d\mathbf{S}d}t} \\ 
	& \text{                          }+\frac{\Delta t}{2{{\tau }^{n}}}\left[ -\Pr \left( 2q_{i}^{n} \right)+{{\Delta }_{i}}2{{\rho }^{n}}R\left( T_{i}^{n}-{{T}^{n}} \right)-2{{\Delta }_{j}}\tau _{ij}^{n}-2{{\Delta }_{k}}\tau _{ik}^{n} \right] \\ 
\end{aligned}
\end{equation}
As can be seen from the above equation, the evolutionary renewal of heat flow depends on the conserved physical quantity at step n, stress, heat flow, and conserved physical quantity at step n+1.\par
At this point, the evolution equation of the distribution function Eq.(\ref{equ3.10}), the evolution equation of the conserved physical quantity Eq.(\ref{equ.3.11}), and the evolution equation of the non-conserved physical quantity \ref{equ3.22} only need to construct ${{f}_{cf}}\left( \mathbf{\xi },t \right)$. When the ${{g}^{*}}$ and $\tau $ of the governing equationEq.(\ref{equ3.1}) are considered as known functions, the governing equation is a linear equation of the distribution function and therefore has an analytical solution, whose expression is
\begin{equation}
{{f}_{cf}}\left( t,\mathbf{\xi } \right)=\frac{1}{\tau }\int\limits_{{{t}_{n}}}^{t}{{{g}^{*}}}\left( {t}',\mathbf{\xi },{{\mathbf{x}}_{cf}}-\mathbf{\xi }t+\mathbf{\xi }{t}' \right){{e}^{\frac{{t}'-t}{\tau }}}d{t}'+{{e}^{\frac{{{t}_{n}}-t}{\tau }}}f\left( {{t}_{n}},\mathbf{\xi },{{\mathbf{x}}_{cf}}-\mathbf{\xi }\left( t-{{t}_{n}} \right) \right)
\end{equation}
This is called local product decomposition. The $f\left( {{t}_{n}},\mathbf{\xi },{{\mathbf{x}}_{cf}}-\mathbf{\xi }\left( t-{{t}_{n}} \right) \right)$ and ${{g}^{*}}\left( {t}',\mathbf{\xi },{{\mathbf{x}}_{cf}}-\mathbf{\xi }t+\mathbf{\xi }{t}' \right)$ of the local product decomposition needs to be constructed by the distribution function and macroscopic physical quantity information at step n.\par
Taking the second-order UGKS method as an example, the distribution function on the interface can be constructed from the distribution function of the cell center and its gradient, expressed as:
\begin{equation}
f\left( {{t}_{n}},\mathbf{\xi },{{\mathbf{x}}_{cf}} \right)=f\left( {{t}_{n}},\mathbf{\xi },{{\mathbf{x}}_{cc}} \right)+{{\left. \frac{\partial f}{\partial \mathbf{x}} \right|}_{cc}}\cdot \left( {{\mathbf{x}}_{cf}}-{{\mathbf{x}}_{cc}} \right)
\end{equation}
The subscript "cc" indicates the cell center,$f\left( {{t}_{n}},\mathbf{\xi },{{\mathbf{x}}_{cf}}-\mathbf{\xi }\left( t-{{t}_{n}} \right) \right)$ can be expressed as
\begin{equation}
f\left( {{t}_{n}},\mathbf{\xi },{{\mathbf{x}}_{cf}}-\mathbf{\xi }\left( t-{{t}_{n}} \right) \right)=f\left( {{t}_{n}},\mathbf{\xi },{{\mathbf{x}}_{cc}} \right)+{{\left. \frac{\partial f}{\partial \mathbf{x}} \right|}_{cc}}\cdot \left( {{\mathbf{x}}_{cf}}-{{\mathbf{x}}_{cc}}+\mathbf{\xi }\left( {{t}_{n}}-t \right) \right)
\end{equation}
The macroscopic physical quantity at the interface can be obtained by calculating the moment of the distribution function already constructed at the interface.

\begin{equation}
{{\mathbf{W}}_{cf}}=\int{\mathbf{\psi }f\left( {{t}_{n}},\mathbf{\xi },{{\mathbf{x}}_{cf}} \right)d\mathbf{\Xi }}
\end{equation}
The equilibrium distribution function ${{g}^{*}}\left( {{t}_{n}},\mathbf{\xi },{{\mathbf{x}}_{cf}} \right)$ at the interface of a unit at time n can be obtained by macroscopic quantities ${{g}^{*}}\left( {t}',\mathbf{\xi },{{\mathbf{x}}_{cf}}-\mathbf{\xi }t+\mathbf{\xi }{t}' \right)$ can be constructed by the following expansion:k

\begin{equation}
\int{\mathbf{\psi }\left( \frac{\partial f}{\partial t}+\mathbf{\xi }\cdot \frac{\partial f}{\partial \mathbf{x}} \right)}d\Xi =0\
\end{equation}
The gradient is obtained by interpolation, while the time derivative is estimated by consistent conditions. The moment of the collision term with respect to the conserved quantity is 0:

\begin{equation}
	\int{\mathbf{\psi }\left( \frac{\partial g}{\partial t}+\mathbf{\xi }\cdot \frac{\partial g}{\partial \mathbf{x}} \right)}d\Xi =0\
\end{equation}
The above formula holds for all distribution functions, and therefore also for equilibrium distribution functions
\begin{equation}
\frac{\partial \mathbf{W}}{\partial t}=-\int{\mathbf{\psi \xi }\cdot \frac{\partial g}{\partial \mathbf{x}}}d\Xi 
\end{equation}
Obtain by integration
Since the gradient of equilibrium distribution function at the interface has been constructed, the time derivative of macroscopic quantity at the interface can be obtained through the above formula, and the time derivative of equilibrium distribution function at the interface can be directly obtained.\par
So far, all terms in the evolution equation of microscopic distribution function and macroscopic physical quantity have been represented by discrete information. The UGKS method is described above from the idea of algorithm construction. The UGKS method is summarized as follows according to the execution steps.\par
(1) Discrete time, physical space and molecular velocity space;\par
(2) The distribution function at the interface of discrete physical units is represented by the local product decomposition of the model equation;\par
(3) In product decomposition, the initial distribution function and the equilibrium distribution function in the molecular motion path are constructed using discrete distribution functions and macroscopic quantities;\par
(4) Using the finite volume idea, the finite volume scheme of the governing equation is constructed, and the evolution equation of the distribution function and macroscopic quantity is obtained;\par
(5) The local volume decomposition at the unit interface is applied to the flux term in finite volume format;\par
(6) The macroscopic quantity of the next time step is obtained by the macroscopic quantity evolution equation;\par
(7) The equilibrium distribution function of the next time step is determined by using the macroscopic quantity of the next time step obtained;\par
(8) The distribution function of the next time step is obtained through the distribution function evolution equation.\par	
	\subsection{Program verification}
\label{sec2.3}
In order to verify the accuracy of the gas dynamics method in the rarefied basin and the continuous basin, the classical high-speed rarefied plate boundary layer flow and laminar plate boundary layer flow were simulated respectively. \par
The numerical example of the flow around the thin plate at high speed is taken from the experimental and numerical simulation of the flow around the thin hypersonic peak front of Tsuboi and Matsumoto. The purpose of the numerical example is to investigate the non-equilibrium phenomenon caused by shock wave-boundary layer interference in the thin hypersonic flow, and to reveal the thermal non-equilibrium variation law of the translational/rotational temperature of diatomic molecular gas. The working gas is nitrogen, and the flow conditions are shown in Table 3-1. The molecular free path in the table is calculated using the VHS molecular model. The values of viscosity index $\omega $, reference viscosity ${{\mu }_{ref}}$, and corresponding reference temperature ${{T}_{ref}}$ in the calculation are set as $\omega =0.75$, ${{\mu }_{ref}}=1.656\times 10-5N\cdot s/{{m}^{2}}$,
${{T}_{ref}}=\text{ }273.15K$. The rotational collision number of Rykov model is set as Z=3.5. The specific configuration and physical space grid dispersion of the slab are shown in Figure 1(a). The slab has a 30° sharp front and a thickness of 15mm. In this work, unstructured grids are used for physical space dispersion Fully thermally adapted diffuse reflection boundary conditions are applied to the surface. The velocity space dispersion is also divided into unstructured grids, as shown in Figure 2(b). The grids around the macroscopic gas velocity of free flow at high speed and low temperature are encrypted, and the total number of velocity space dispersion points is 2838.\par
Figure 3 shows the flatbed translational temperature cloud image and rotation temperature cloud image obtained by the gas dynamics method using the flow reference to perform dimensionless calculation. In hypersonic rarefied flow, the most obvious feature is the non-equilibrium effect. Whether the non-equilibrium effect can be accurately captured is a good verification for the accuracy of the algorithm The peak value of the leading edge is reached, and the peak value of the rotational temperature is reached at the lower part of the leading edge of the plate, which indicates the conversion process between the translational temperature and the rotational temperature, shows the non-equilibrium characteristics of the leading edge of the plate, and shows the thin simulation ability of the method. The comparison of dimensional calculation results with DSMC method and experimental results is shown in Figure 4. It can be seen that there is a large non-equilibrium phenomenon between the translational and rotational temperature of the flow near 5mm of the front edge of the plate, while the phenomenon is weakened at 20mm, indicating that the rotational and dynamic temperatures gradually tend to be balanced with the development of the flow. The rotational temperature obtained by this method is in good agreement with the experimental results.

Then, the algorithm accuracy of the gas dynamic method in a continuous basin is verified. Laminar plate boundary layer flow is a classic example to test the accuracy of the method to simulate boundary layer flow. In this example, the incoming Mach number is 0.16 and the Reynolds number is 100,000, which corresponds to a reference length of 100mm plate length. The height of the first layer grid is 0.02,as shown in Figure 5. The adiabatic wall of the plate is taken, the symmetric boundary conditions (x<0,y=0) are taken in front of the plate, and the subsonic inlet boundary conditions are taken in the other three directions. The velocity grid in the range [-4,4] is discretized to 101×101 Simpson integral points.\par

The reference result is the Blasius solution. In order to better verify the accuracy of the method, the velocity patterns at 29mm, 50mm and 70mm of the plate are respectively taken for comprehensive comparison, as shown in Figure 6. It can be seen that the simulation results of the gas kinetics method are in good agreement with the Blasius solution at different positions of the plate. The simulation capability of gas kinetics method in continuous watershed is proved.\par
	\section{Numerical results and analysis}
\label{sec3}
\subsection{Two-dimensional corrugated wall plate simulation}~{}
\label{sec3.1}
According to the flow field obtained by the simulation of the slab example, the corrugated wall is arranged. The distribution range of the corrugated wall layout is near the synchronization point of the least stable mode. The corrugated wall profile adopts sinusoidal configuration, and the generation function of the normal coordinate of the wall is
		\begin{equation}\label{equ3.35}
y={{A}_{w}}sin(\frac{2\pi }{{{\lambda }_{x}}}x+\frac{\pi }{2})-{{A}_{w}}
		\end{equation}
where A is the amplitude of the corrugated wall shape function and 2 is the wavelength of the unstable second mode flow direction.In this example, the layout range of the corrugated wall is set to 10 times the wavelength (10${{\lambda }_{x}}$), and three kinds of corrugated walls with different heights are selected for study. The height of the three corrugated walls ranges from small to large 0.2${{\delta }_{0}}$ respectively., 0.4${{\delta }_{0}}$, 0.6${{\delta }_{0}}$,(${{\delta }_{0}}$ is the local boundary layer thickness) the type (3.35),${{A}_{w}}=0.1{{\delta }_{0}},0.2{{\delta }_{0}},0.3{{\delta }_{0}}$, and three examples respectively for W1, W2, W3, the flow condition is the same as that shown in Table 3-3. The calculation domain is that the flow direction position ranges from -50-500, the normal calculation domain height is 50, and the supersonic inlet and outlet conditions are adopted at the entrance and exit. The wall surface adopts diffuse reflection boundary condition. The upper boundary adopts the inflow condition outside the boundary layer, and the calculation mesh adopts body-fitted mesh, and is encrypted near the corrugated wall. The corrugated wall ensures that there are at least 30 grid points in each wavelength. The height of the first layer grid is 0.001, the number of normal grid points in the boundary layer is about 81, and the total number of grid points is 151×751=113401, and the inflow reference is used Dimensionless.\par
Figure 7 shows the flow diagram near the corrugated wall. Because the layout of the corrugated wall is a groove-like structure below the wall, a certain vortex structure is formed in the corrugated wall. Figure 8 shows that the vortex volume increases to a certain extent with the increase of the depth of the corrugated wall. The emergence of the vortex structure will generate a certain heating phenomenon for the flow near the corrugated wall. Therefore, the temperature of the boundary layer near the corrugated wall will be higher than that at the smooth wall surface, as shown in Figure 9. It can be seen that the existence of the corrugated wall has a certain deceleration effect on the flow near the wall, resulting in a stronger heating phenomenon in the flow near the corrugated wall. \par
Since the corrugated wall is a tiny structure relative to the entire plate in a single wavelength, there should be some discussion about whether there is a certain thinning effect inside the corrugated wall. We should note that in the simulation of the non-abrupt plate, we have reached a conclusion that the continuity of the flow will be strengthened as the flow develops along the plate, so if there is no thinning effect in the front, the flow will be strengthened. At the same time, it is noted that the flow Reynolds number of this calculation is very large, which is a very continuous calculation example, and the arrangement range of the corrugated wall belongs to the middle and back position of the plate, so the front part of the junction between the corrugated wall and the plate will not appear local thinning phenomenon, so our observation point is inside the corrugated wall. Taking the first wavelength as the observation object, the positions of wave crest X=304 and trough X=308 were tested respectively. Figure 10~12 show the dimensionless translational and rotational temperature cloud maps of the flow temperature at the first crest trough of three different corrugated walls. One of the most obvious features is that in the case of thin flow, the collision frequency between molecules is less, resulting in the non-equilibrium phenomenon of translational and rotational temperatures of different degrees of freedom. For two-dimensional flow, it is the inequality of rotational and dynamic temperatures. It can be seen from the extracted profile line that there is no inequality at either wave crest or trough. It shows that the flow at the corrugated wall is a continuous flow with no non-equilibrium effect for the example of this working condition.\par
Figure 13~14 shows the dimensionless pressure and shear force of the wall near the corrugated wall. It can be seen from the pressure curve that, on the whole, the pressure at the corrugated wall changes little compared with the incoming flow, even at the maximum peak value, it does not exceed 1.2 times of the incoming flow, except that the pressure near the first wavelength is lower and the pressure near the last wavelength is higher. The pressure of other wavelengths is more equal, and the pressure, shear stress and heat flow peak also increase with the increase of the depth of the corrugated wall. In addition, the reason why the peak pressure at the first corrugated wall is higher and the peak value at the last corrugated wall is lower is that the flow at the first wavelength is hindered and then flows downstream rapidly, resulting in a lower peak pressure here. At the last wavelength, because the downstream is a smooth wall, the flow does not have a downstream dredding effect as at the previous wavelength. In contrast, the flow is blocked to a certain extent, resulting in increased pressure. As for the shear force, the reason why it fluctuates within one wavelength is due to the influence of the strength of the flowing vortex structure, while the position of the peak value is related to the position of the vortex structure. The peak value within the wavelength moves backward and its size increases with the increase of the depth of the corrugated wall, which indicates that the vortex structure is strengthened with the deepening of the corrugated wall. Figure 15 shows the dimensionless heat flow at the corrugated wall, and Figure 16 shows the thickness of the temperature boundary layer near the corrugated wall. The thickness of the temperature boundary layer is defined by$(T-{{T}_{wall}})=0.99({{T}_{in}}-{{T}_{wall}})$ .Note that the size of the wall heat flow is related to the size of the temperature boundary layer, and the flow field heat flow in the thick temperature boundary layer is small, while the flow field heat flow in the thin boundary layer is large. Due to the existence of the internal vortex structure, the thickness of the temperature boundary layer is small on both sides and large in the middle. The surface heat flow oscillates at the corrugated wall and reaches the maximum value at the trough.\par
In order to study the influence of rarefied effect on flow, flow simulation was performed again after the decrease of incoming flow Reynolds number. Reynolds number ${{10}^{6}}$, ${{10}^{5}}$, ${{10}^{4}}$, corresponding heights of about 33km, 50km and 68km were respectively used to simulate the corrugated wall flow with a height of 0.6. The wall surface and flow field parameters at the corrugated wall were mainly observed. It should be noted that as the inflow Reynolds number decreases, the thickness of the boundary layer increases, and the mesh height of the first layer needs to be increased to 0.001mm, 0.005mm, 0.01mm, 0.05mm, respectively. As shown in Figure 17, the velocity patterns of translational and rotational temperatures at the first trough of the corrugated plate wall are extracted. It can be found that when the incoming flow Reynolds number drops to ${{10}^{5}}$, the non-equilibrium phenomenon of translational temperature and rotational temperature begins to appear at the corrugated wall, and when the incoming flow Reynolds number drops to ${{10}^{4}}$, this thermal non-equilibrium effect has become very significant.\par
Figure 18 shows the local Knudsen number near the corrugated wall. The continuous failure criterion proposed by Boyd et al., $K{{n}_{GLL}}=\frac{\lambda}{Q/\nabla Q}$(where is the molecular mean free path, the VHS model is used for calculation in this simulation, and $Q$refers to the main flow variable. In this criterion, when the local Knudsen number is greater than 0.05, it can be considered that the continuity hypothesis fails and the flow becomes thin flow. It can be found that when the flow passes through the corrugated wall, its local Knudsen number always increases first and then decreases, because when the flow passes through the corrugated wall, its local Knudsen number always decreases. Because the corrugated wall is a concave structure, the flow into the corrugated wall will produce a certain expansion, resulting in the increase of Knudsen number. For the windward side of the corrugated wall, when the flow passes through the bottom to the windward side, it will produce a certain compression effect, which reduces the average molecular free path of gas molecules, resulting in the reduction of Knudsen number. At the same time, with the decreasing of the Reynolds number, the local Knudsen number of the corrugated wall also increases. On the other hand, it can be seen that when the Reynolds number of incoming flow is ${{10}^{7}}$,${{10}^{6}}$, the local Knudsen number near the corrugated wall has not reached the condition of continuous failure, and there is no non-equilibrium effect in the temperature profile at this time. When the inflow Reynolds number reaches ${{10}^{5}}$, the peak Knudsen number of the local corrugated wall has reached 0.05, that is, the continuous failure criterion is satisfied, and at this time, the thermal non-equilibrium effect of translational and rotational temperatures also begins to appear, which also verifies the accuracy of the theory from another hand. When the inflow Reynolds number is further reduced to ${{10}^{4}}$, the local Knudsen number is larger, and the non-equilibrium phenomenon of translational temperature and rotational temperature has been very serious. It can be seen that for the model and working conditions, the inflow Reynolds number of ${{10}^{5}}$ or the height of 50km is a critical value. When the height is lower than 50km, it can be considered that the influence of the corrugated wall on the flow rarefaction degree is not needed to be considered, while when the flight height is higher than 50km, the influence of rarefaction effect on the flow needs to be considered to some extent.\par
	\begin{table}
	\begin{tabular}{cc}
		\hline
		flow variables & variable value\\
		\hline
		Freestream temperature ${{T}_{\infty }}$, K & 79 \\ 
		Prandtl number& 0.72 \\
		specific heat ratio & 1.4\\
		 Freestream Reynolds number	, mm-1& ${{10}^{7}}$,${{10}^{6}}$,${{10}^{5}}$,${{10}^{4}}$\\
      	Freestream Mach number & 6 \\
		Plate temperature ${{T}_{\text{wall}}}$, K & 294 \\
		\hline
	\end{tabular}
	\centering 
	\caption{\label{table3.1}Computational conditions for thin hypersonic flow around a plate.}
\end{table}
	\subsection{Two-dimensional sharp wedge simulation}~{}
	\label{sec3.2}
	In order to simulate more condition, the two-dimensional sharp wedge is used to simulate the flow, and the calculation conditions are shown in the following table. For the sharp wedge with abrupt change, the generation function of the corrugated wall is consistent with that of the flat plate. In this example, the arrangement range of the corrugated wall is set as 10 times the wavelength (10${{\lambda}_{x}}$), and the synchronization position is 60mm at the sharp wedge. Corrugated wall height is 0.4 ${{\ delta} _ {0}} $(${{\ delta} _ {0}} $for local boundary layer thickness), namely the type (3.35) in the ${{A} _ {w}} = 0.2 {{\ delta} _ {0}} $. Figure 19 shows the computing grid. The flow direction position ranges from -50 to 500, the normal direction height is 150, and the supersonic inlet and outlet conditions are adopted. The wall surface adopts diffuse reflection boundary condition. The upper boundary adopts the inflow condition outside the boundary layer, and the calculation mesh adopts the body-fitted mesh, which is encrypted near the corrugated wall. The corrugated wall ensures that there are at least 30 mesh points in each wavelength. The height of the grid in the first layer is 0.001, the number of mesh points in the boundary layer is about 81, and the total number of mesh points is 235500. The velocity space is encrypted at the stationary point velocity and the incoming flow velocity, the number of grids is 2872, and the incoming flow parameters are used for non-dimensionalization. In order to study the influence of rarefication effect on flow, flow simulation was performed again after the Reynolds number of incoming flow was reduced, and Reynolds number ${{10}^{6}}$, ${{10}^{5}}$ and ${{10}^{4}}$ were used respectively, corresponding heights of about 33km, 50km and 68km respectively. It should be noted that as the inflow Reynolds number decreases, the thickness of the boundary layer increases, and the mesh height of the first layer needs to be increased to 0.005mm, 0.01mm and 0.05mm respectively.\par
	
	Figure 20 shows the flow diagram near the corrugated wall, which is similar to the plate flow. Because the corrugated wall is arranged below the wall, it is a groove-like structure. When the Reynolds number of incoming flow is ${{10}^{7}}$ and ${{10}^{6}}$, vortex structure will appear in the structure.\par
	Figure 21 shows the local Knudsen number cloud map near the sharp wedge corrugated wall. Similar to the pattern of flat plate flow, the maximum Knudsen number is reached at the top of the windward and leeward sides, indicating a certain thinning effect caused by flow expansion, while the bottom of the corrugated wall decreases Knudsen number due to the blocking effect of the leeward side.\par
	Figure 22 further shows the specific performance of the local Knudsen number near the corrugated wall gradually increasing in translational and rotational temperatures as the incoming Reynolds number decreases. Through the temperature profile of different incoming Reynolds numbers at the first trough of the corrugated wall (i.e., X=65), we can see that when the incoming Reynolds number is ${{10}^{7}}$ and ${{10}^{6}}$, The temperature profile at the trough does not clearly reflect the non-equilibrium effect of the translational and rotational temperatures, and when the inflow Reynolds number drops to ${{10}^{5}}$, the non-equilibrium effect of the translational and rotational temperatures has begun to appear, which also confirms that the continuous failure criterion has been met as shown in the local Knudsen number cloud map. When the Reynolds number reaches ${{10}^{4}}$, the temperature disequilibrium becomes more severe, and the local Knudsen number also becomes larger.\par
		\begin{table}
		\begin{tabular}{cc}
			\hline
			flow variables & variable value\\
			\hline
			Freestream temperature ${{T}_{\infty }}$, K & 79 \\ 
			Prandtl number& 0.72 \\
			specific heat ratio & 1.4\\
			Freestream Reynolds number	, mm-1& ${{10}^{7}}$,${{10}^{6}}$,${{10}^{5}}$,${{10}^{4}}$\\
			Freestream Mach number & 6 \\
			Plate temperature ${{T}_{\text{wall}}}$, K & 294 \\
			Angle of attack ${{T}_{\text{wall}}}$, $^{\circ }$ & 6 \\
			Semiapex angle ${{T}_{\text{wall}}}$, $^{\circ }$ & 7 \\
		Overall length ${{T}_{\text{wall}}}$, mm & 500 \\
			\hline
		\end{tabular}
		\centering 
		\caption{\label{table3.2}Computational conditions for thin hypersonic flow around a plate.}
	\end{table}

	\section{Conclusion}
	\label{sec4}
	Exploring the microscopic mechanism of micro-mutation has important guiding significance for the aerodynamic optimization design of new hypersonic vehicles in the future, and plays a theoretical basis and database role for the research and development of new hypersonic vehicles.Boundary layer flows have a significant impact on the aerodynamic/thermal performance of near-space hypersonic vehicles, space shuttle vehicles and future space and space vehicles, while the microstructure and roughness of the wall surface have an important impact on boundary layer flows. Therefore, to accurately predict the forces and heat transfer of such cross-basin vehicles, it is necessary to study the microscopic mechanism of micro-sudden changes on the surface by means of a unified whole-basin approach. In this project, DNS method and UGKS method are used to study the rough wall flow mechanism, which provides a new efficient and accurate numerical prediction method for the study of cross-basin rough wall flow. This work has important guiding significance for the aerodynamic optimization design of new hypersonic vehicles in the future, and plays a theoretical basis and rich database role for the research and development of new hypersonic vehicles.
	
	\section*{Acknowledgments}
		The authors would like to thank Junlei Mu, Qingdian Zhang, Boxiao Zou and Peiyuan Geng for helpful discussion.

	\clearpage
	\begin{figure}[H]
		\centering
		\subfloat{\includegraphics[width=1.0\textwidth]{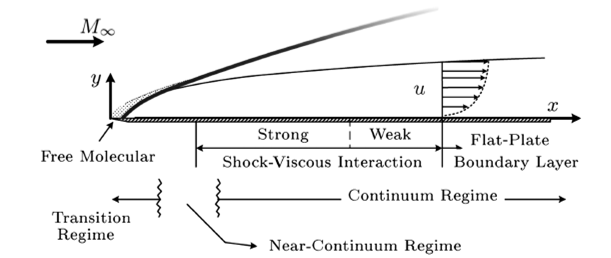}}
		\caption{\label{1-1}Schematic diagram of hypersonic incoming flow around the leading edge flat plate at zero Angle of attack.}
	\end{figure}

	\begin{figure}[H]
		\centering
        \subfloat[Physical mesh]{\includegraphics[width=0.4\textwidth]{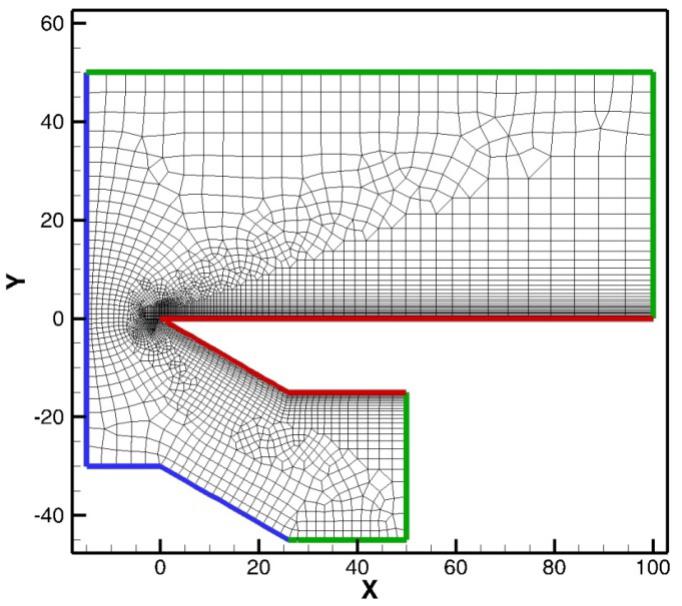}}
		\subfloat[Velocity mesh]{\includegraphics[width=0.4\textwidth]{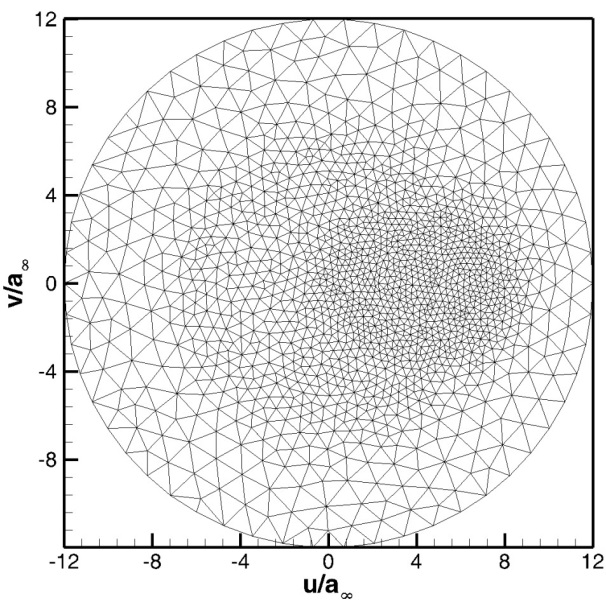}}
		\caption{The grid used for high-speed rarefied flow plate calculation example }
		\label{3-5}
	\end{figure}

\begin{figure}[H]
	\centering
	\subfloat[Trasition temperature]{\includegraphics[width=0.4\textwidth]{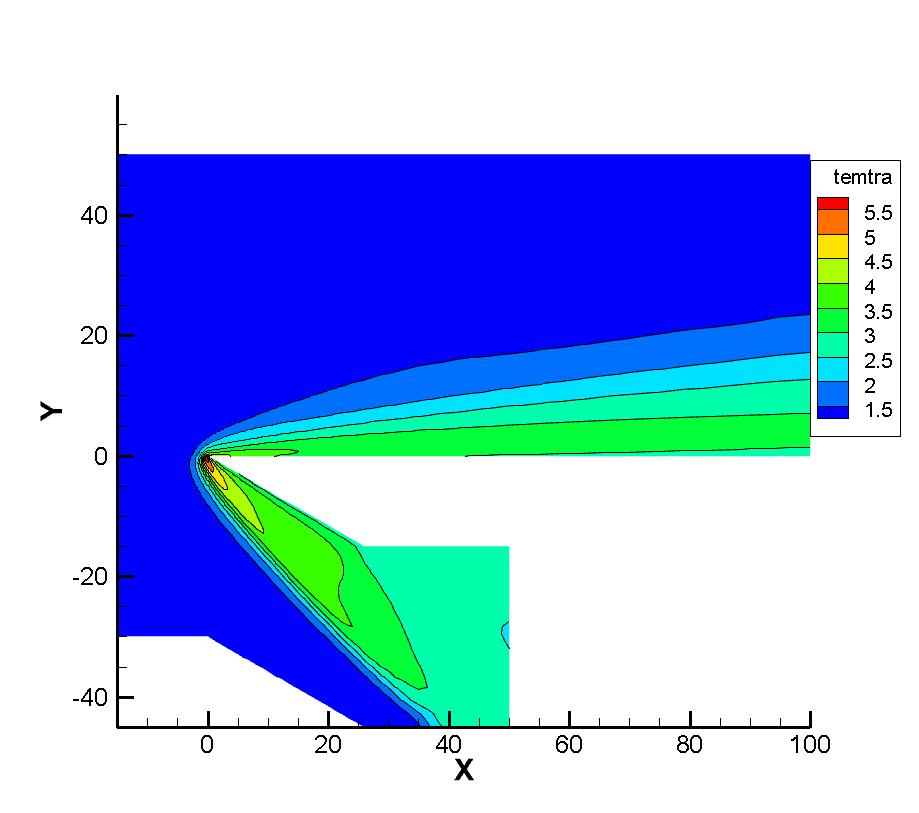}}
	\subfloat[Rotation temperature]{\includegraphics[width=0.4\textwidth]{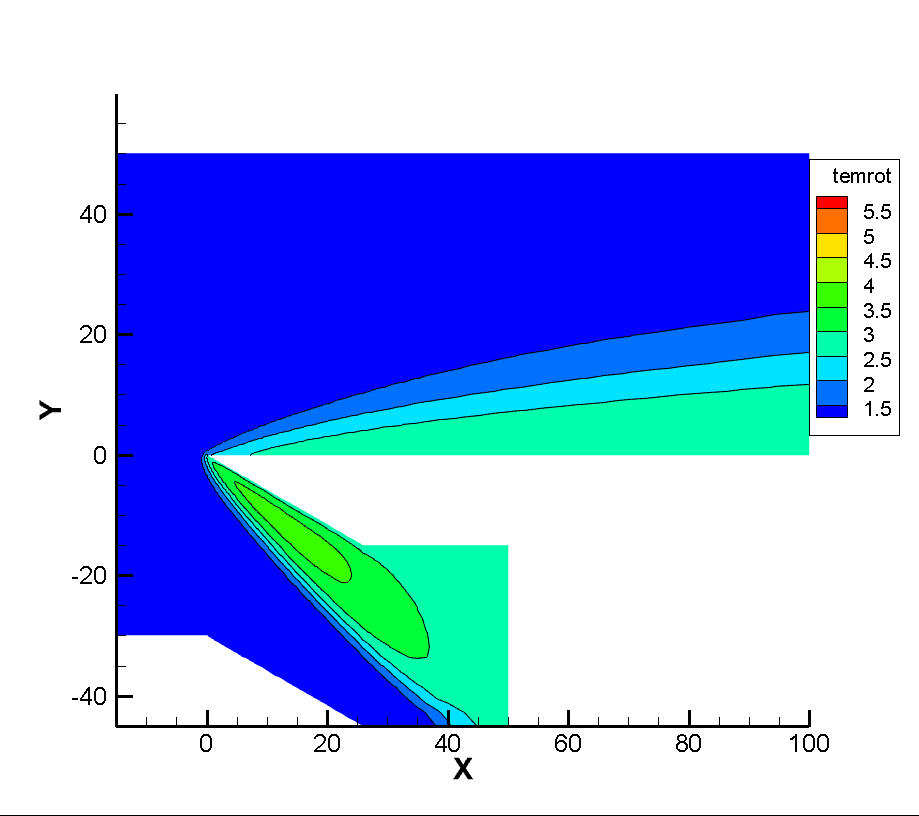}}
	\caption{High-speed rarefied flow plate example translational and rotational temperature cloud image }
	\label{3-6}
\end{figure}

\begin{figure}[H]
	\centering
	\subfloat[X=5mm]{\includegraphics[width=0.4\textwidth]{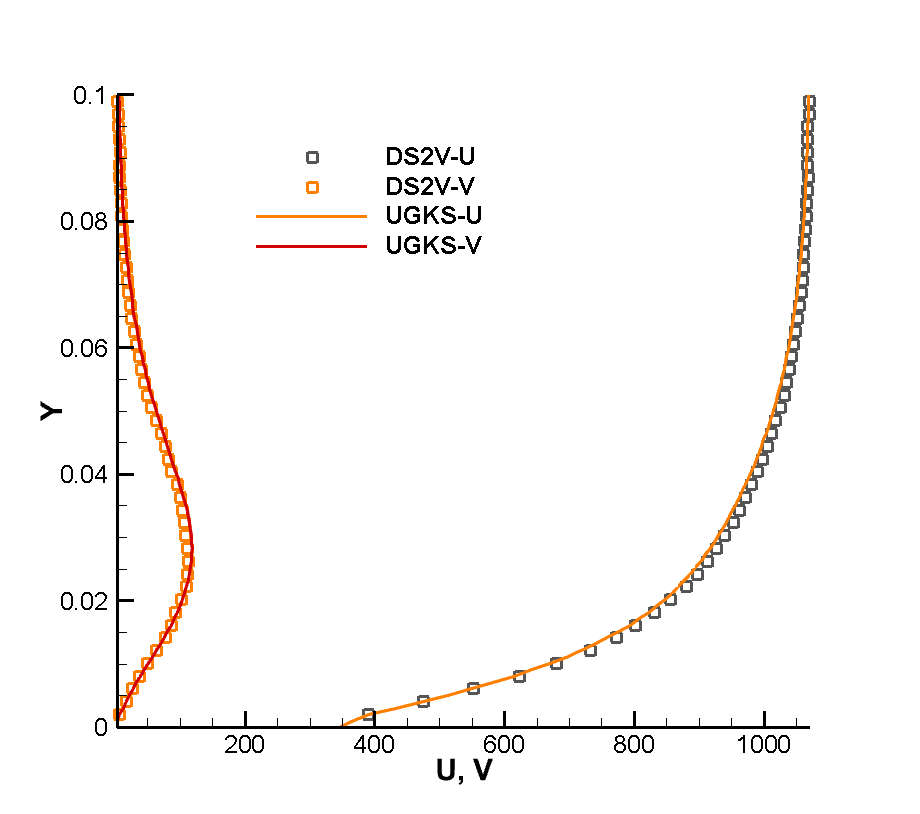}}
	\subfloat[X=20mm]{\includegraphics[width=0.4\textwidth]{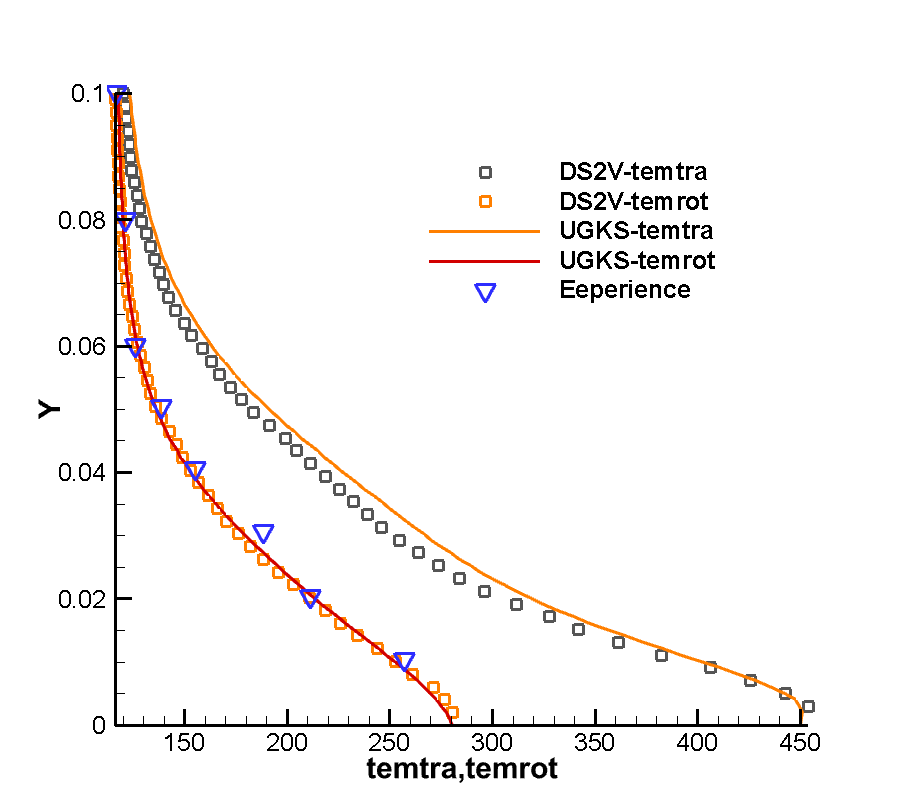}}
	\caption{Comparison of the normal temperature distribution of the flow around a high speed thin plate at different locations with experimental values and DSMC results }
	\label{3-7}
\end{figure}

	\begin{figure}[H]
		\centering
		\includegraphics[width=0.8\textwidth]{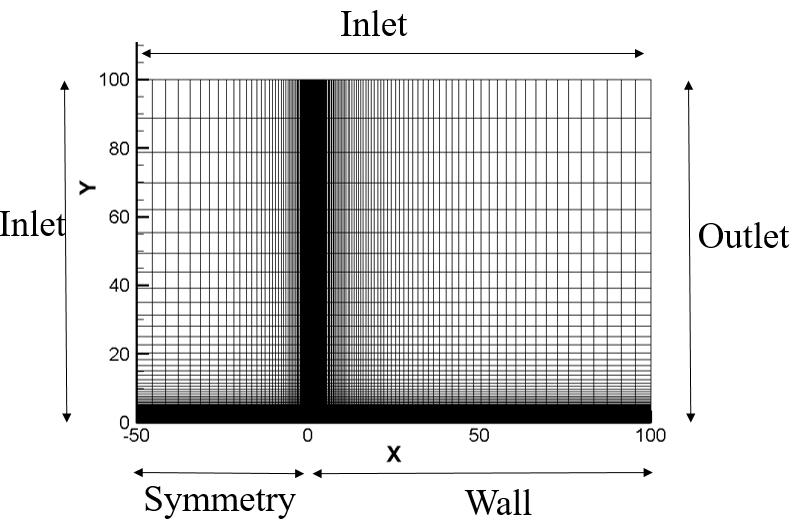}
		\caption{\label{3-8}Laminar flow plate boundary layer mesh (30351 mesh).}
	\end{figure}
	
	\begin{figure}[H]
		\centering
		\subfloat{\includegraphics[width=0.4\textwidth]{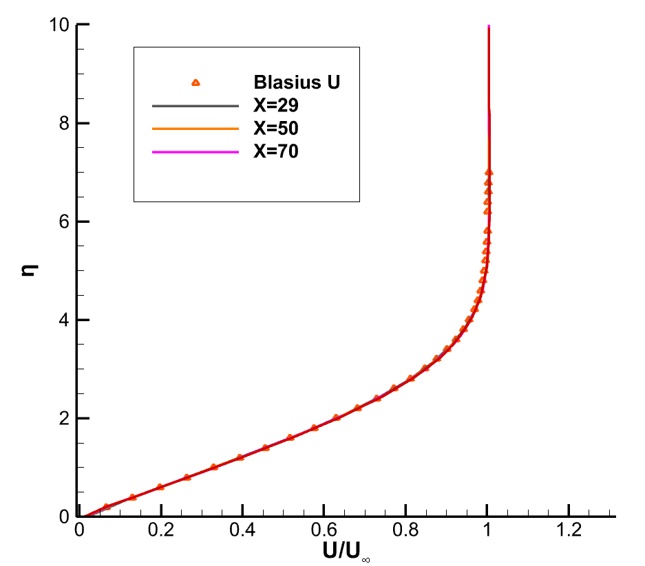}}
		\subfloat{\includegraphics[width=0.4\textwidth]{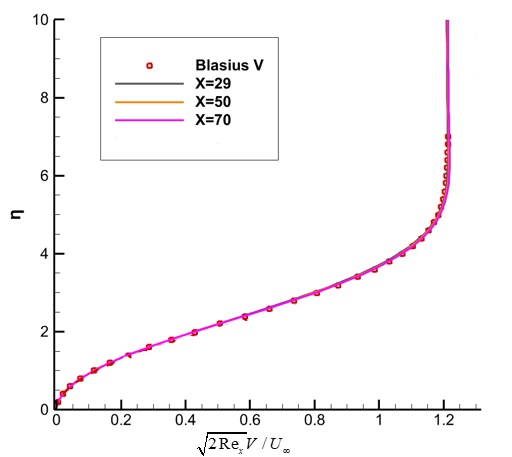}}
		\caption{Laminar plate boundary layer velocity profile}
		\label{3-9}
	\end{figure}

	\begin{figure}[H]
		\centering
		\subfloat[W1]{\includegraphics[width=0.3\textwidth]{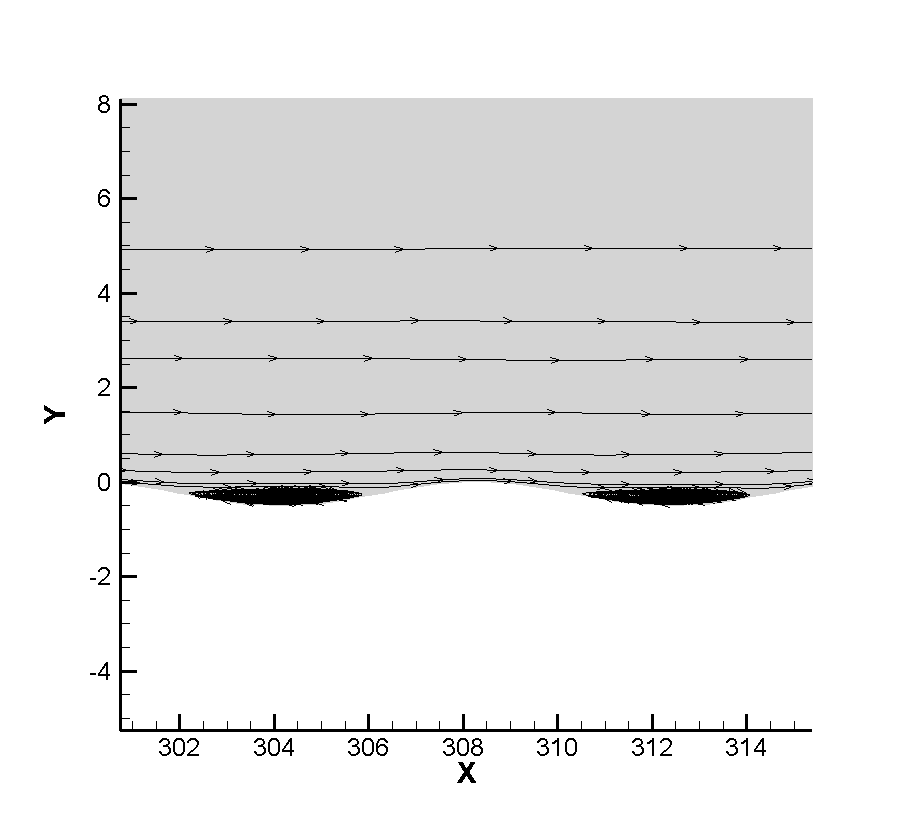}}
		\subfloat[W2]{\includegraphics[width=0.3\textwidth]{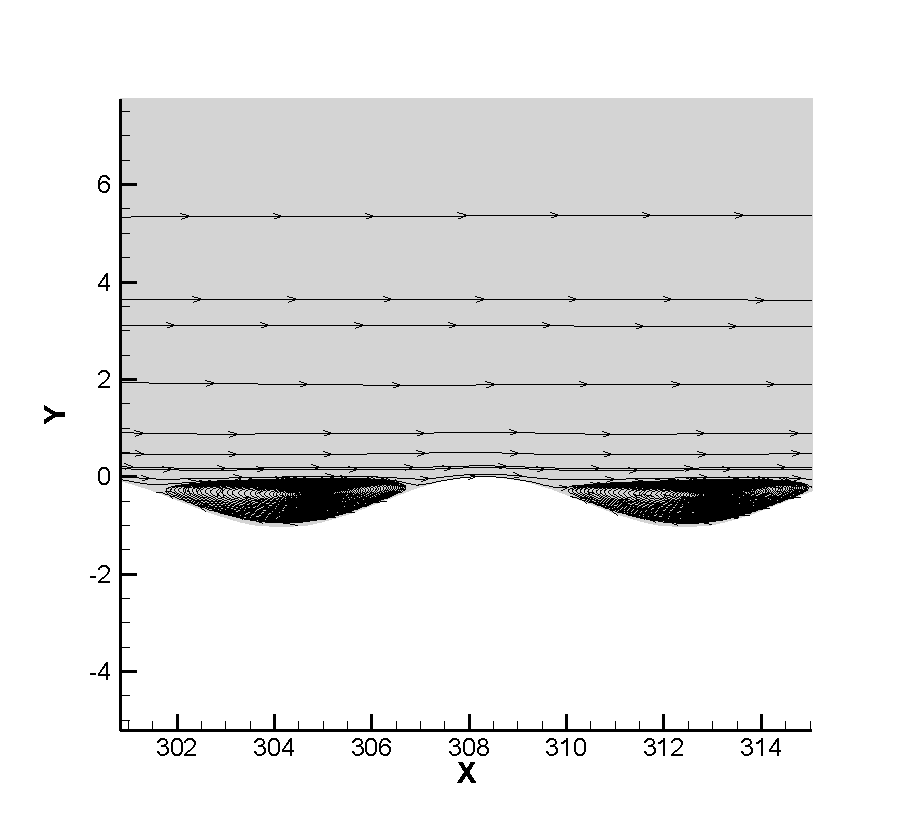}}
		\subfloat[W3]{\includegraphics[width=0.3\textwidth]{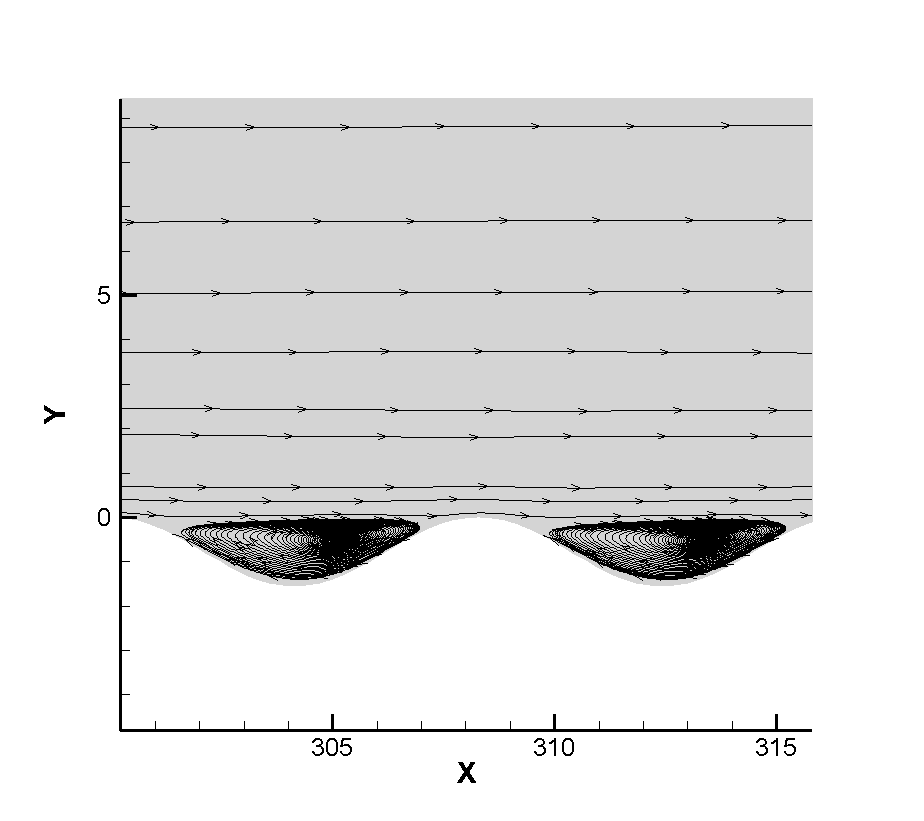}}

		\caption{Flow diagram near the corrugated wall }
		\label{3-15}
	\end{figure}

	\begin{figure}[H]
	\centering
	\subfloat[W1]{\includegraphics[width=0.3\textwidth]{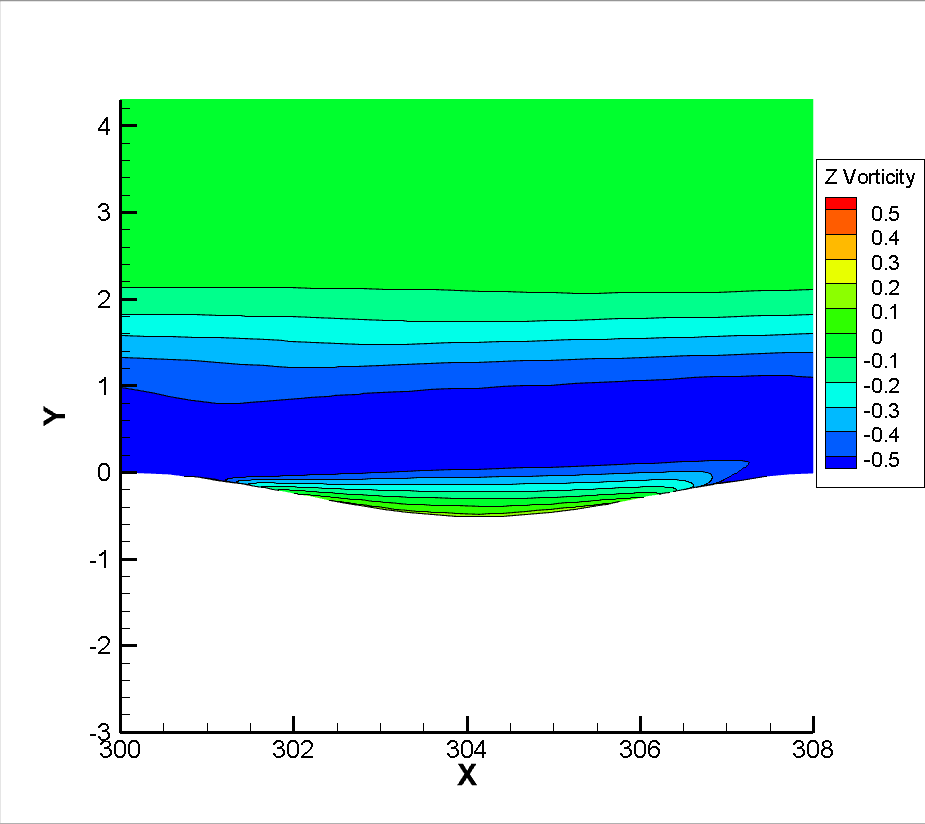}}
	\subfloat[W2]{\includegraphics[width=0.3\textwidth]{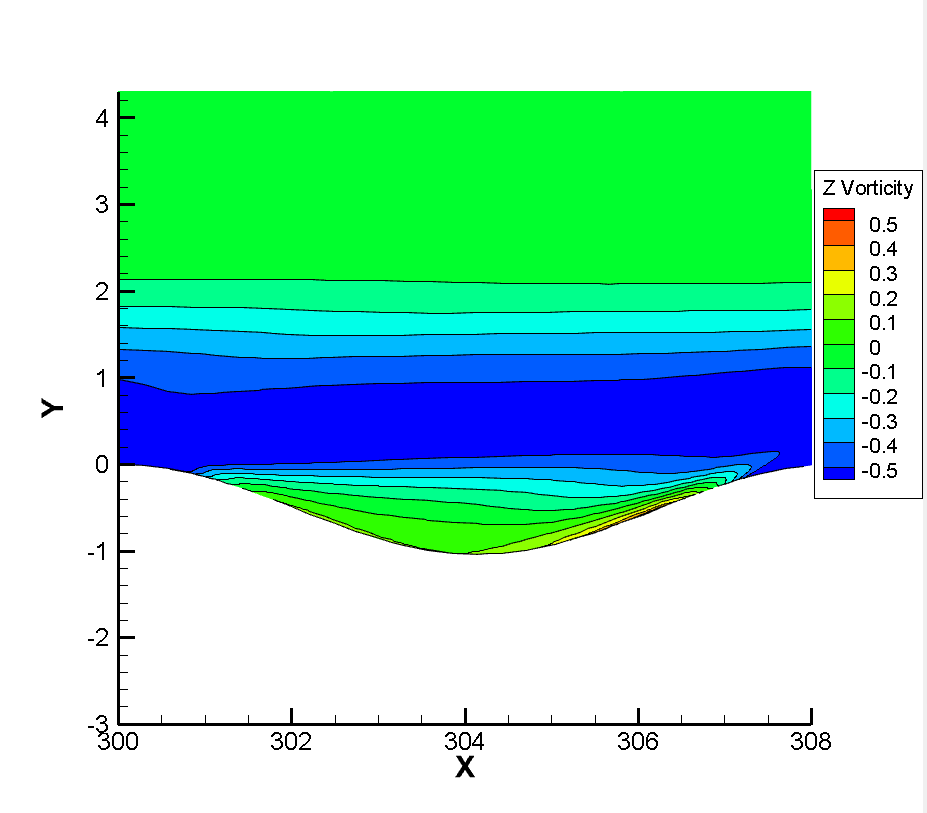}}
	\subfloat[W3]{\includegraphics[width=0.3\textwidth]{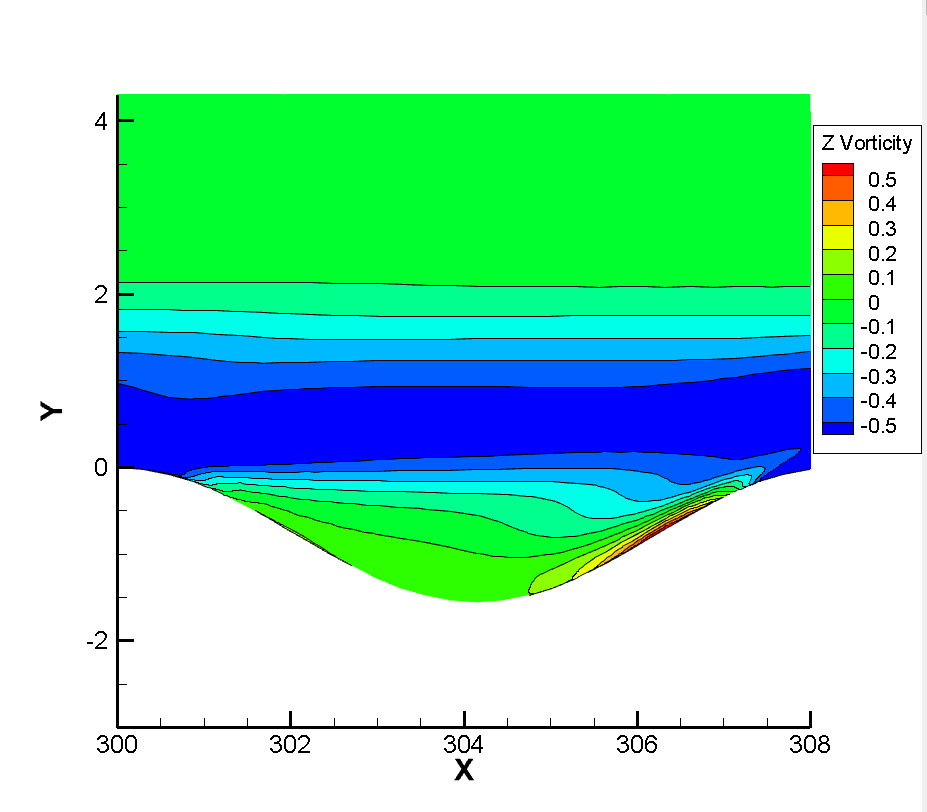}}
	
	\caption{Vorticity diagram near the corrugated wall }
	\label{3-16}
\end{figure}

	\begin{figure}[H]
	\centering
	\subfloat[W1]{\includegraphics[width=0.3\textwidth]{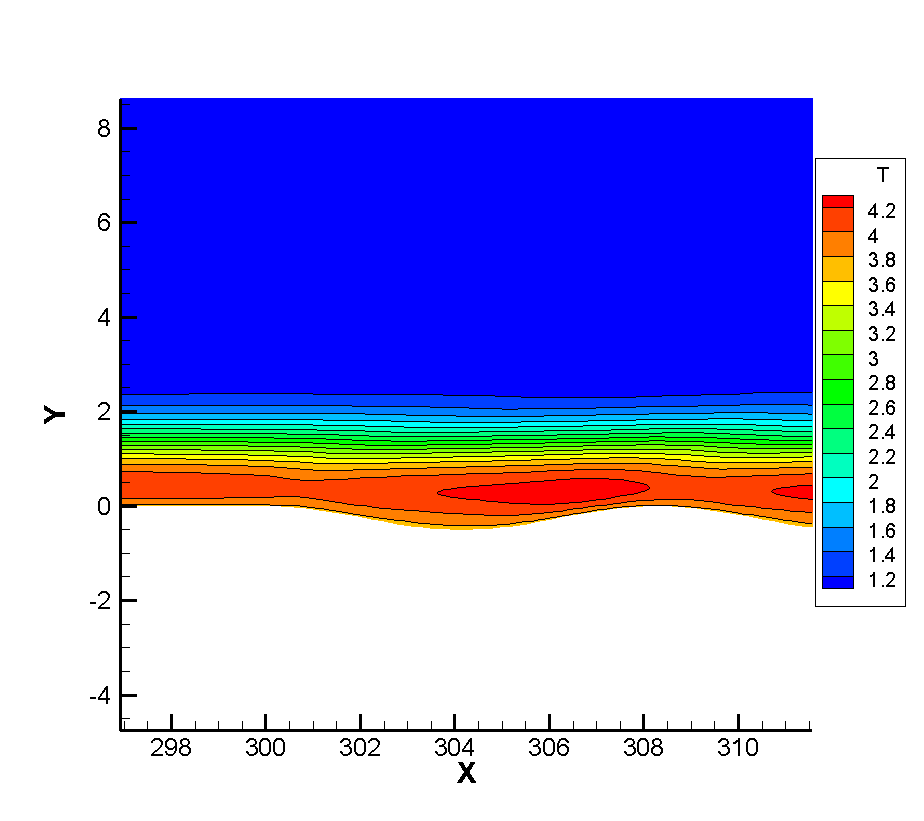}}
	\subfloat[W2]{\includegraphics[width=0.3\textwidth]{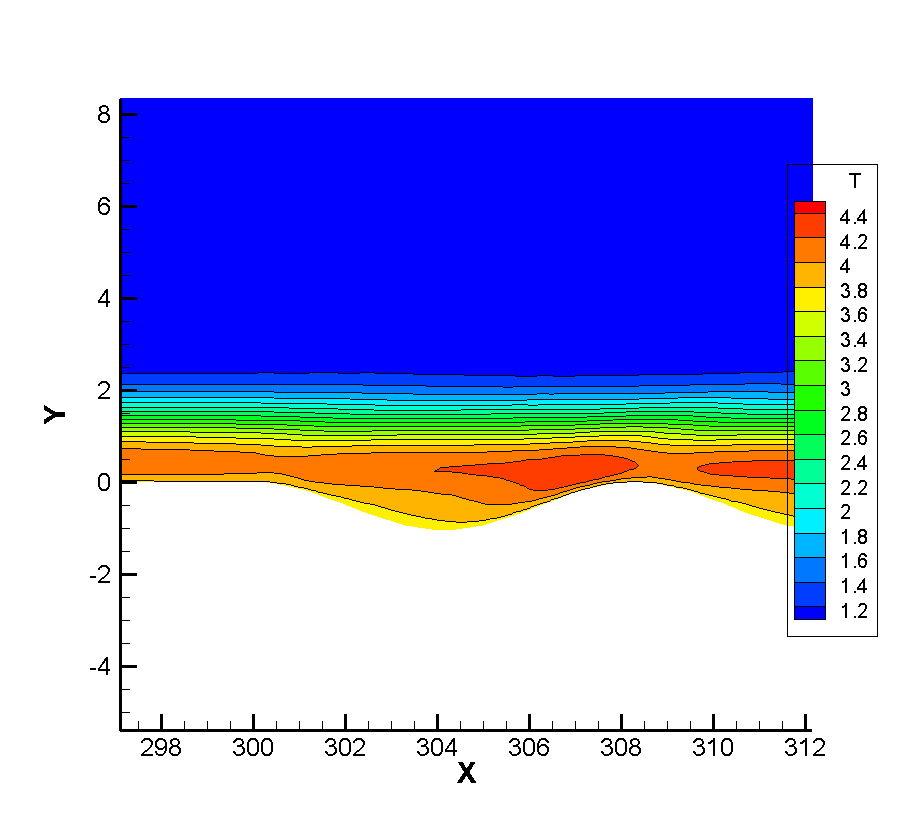}}
	\subfloat[W3]{\includegraphics[width=0.3\textwidth]{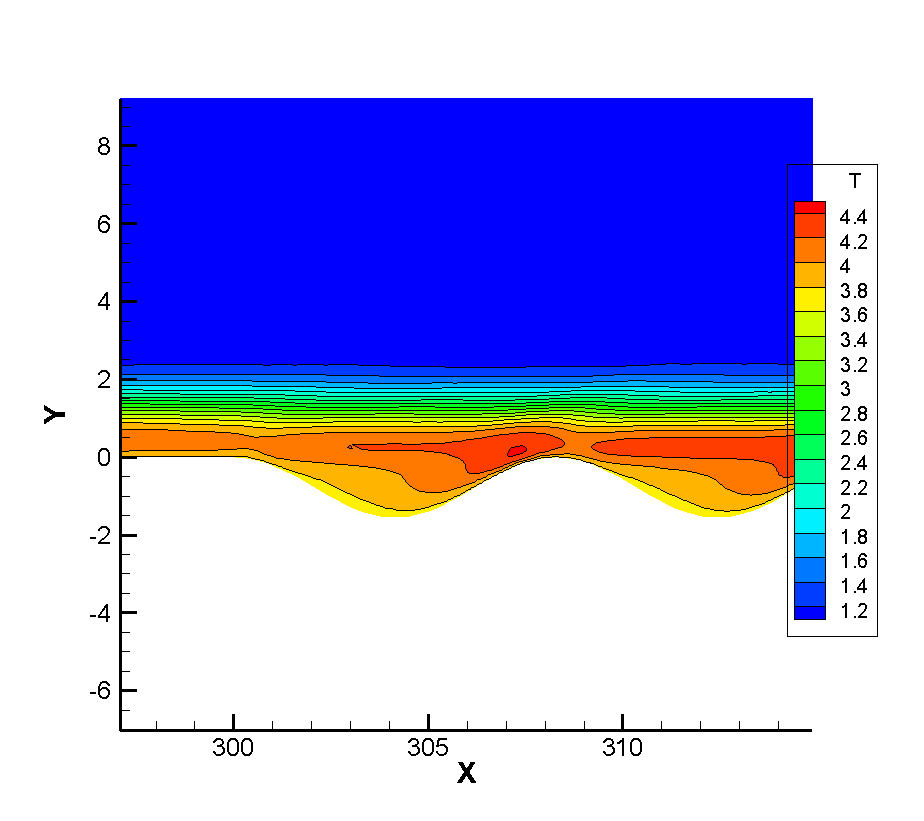}}
	
	\caption{Total temperature cloud image near corrugated wall }
	\label{3-17}
\end{figure}

	\begin{figure}[H]
	\centering
	\subfloat[X=304mm]{\includegraphics[width=0.4\textwidth]{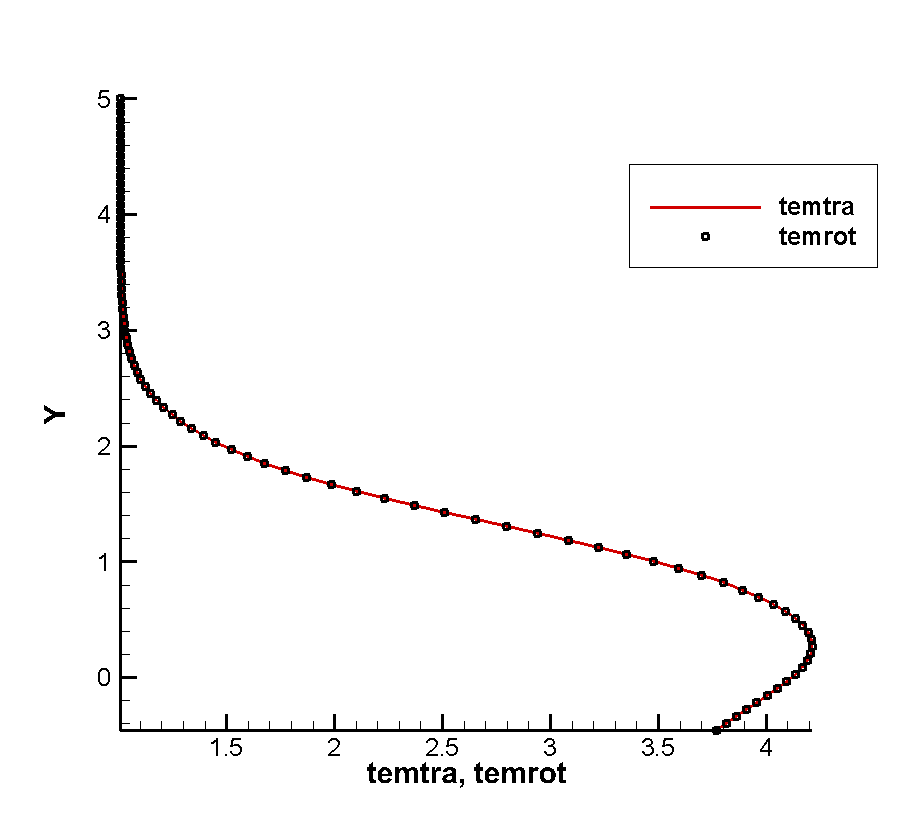}}
	\subfloat[X=308mm]{\includegraphics[width=0.4\textwidth]{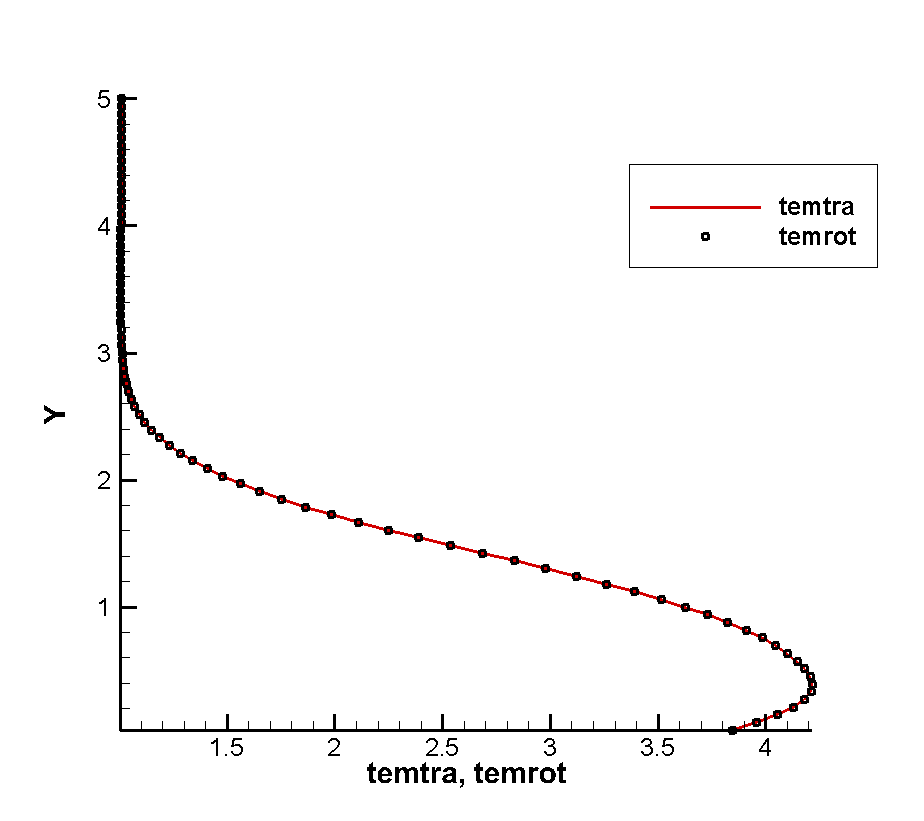}}
	\caption{Translational and rotational temperature profiles at wave peaks and troughs of corrugated wall W1}
	\label{3-18}
\end{figure}

\begin{figure}[H]
	\centering
	\subfloat[X=304mm]{\includegraphics[width=0.4\textwidth]{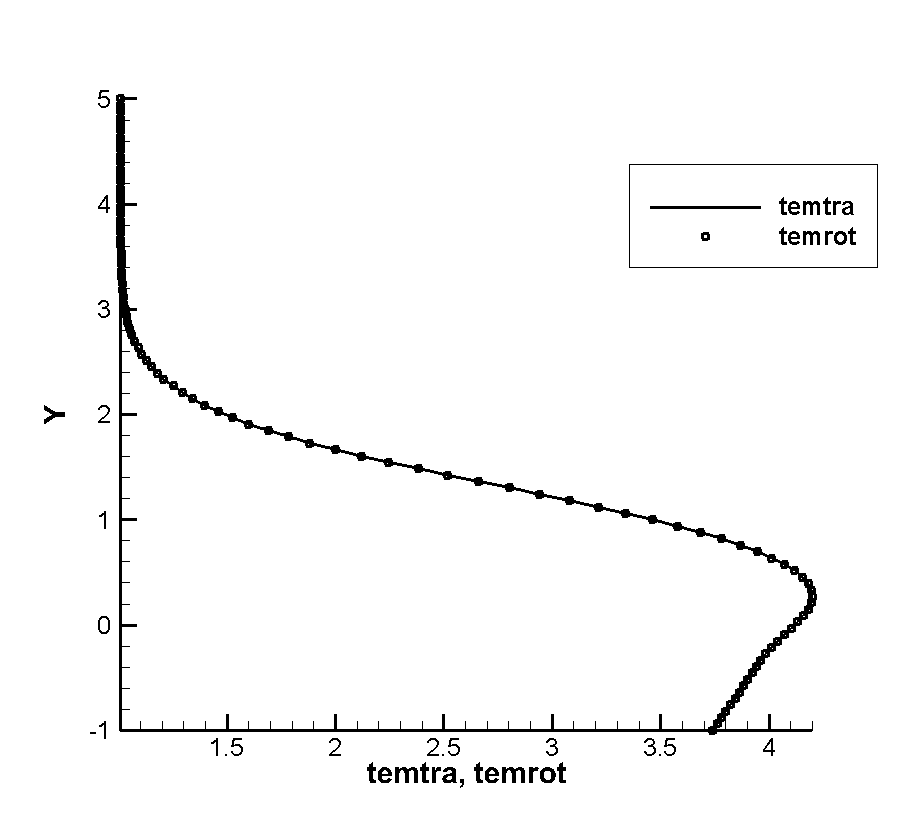}}
	\subfloat[X=308mm]{\includegraphics[width=0.4\textwidth]{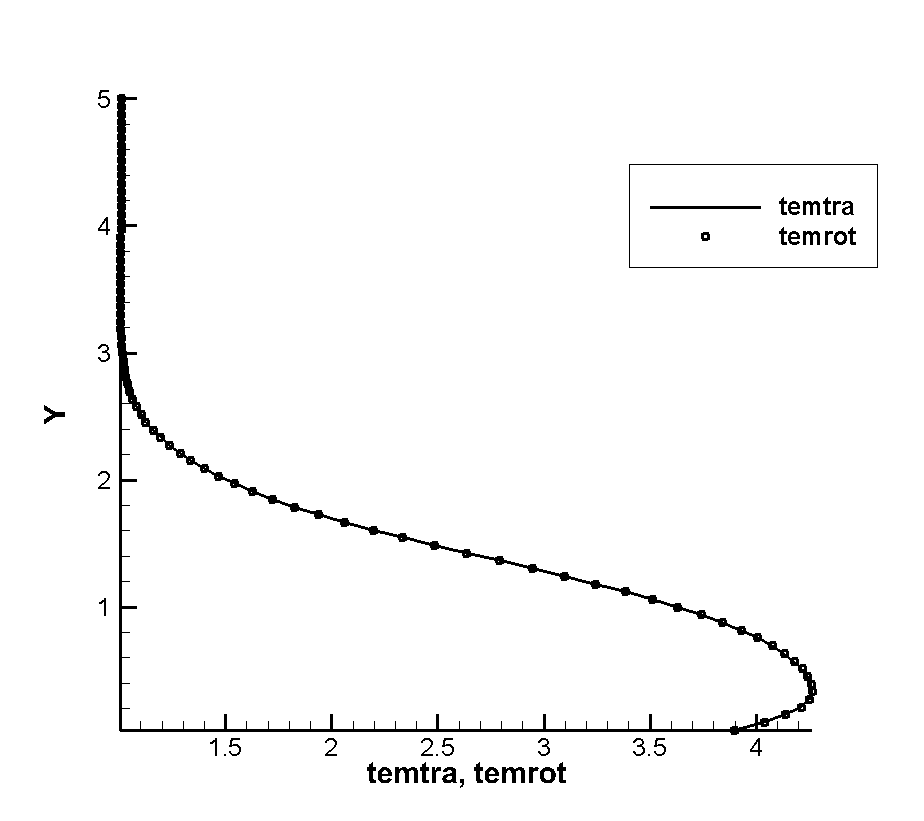}}
	\caption{Translational and rotational temperature profiles at wave peaks and troughs of corrugated wall W2}
	\label{3-19}
\end{figure}

\begin{figure}[H]
	\centering
	\subfloat[X=304mm]{\includegraphics[width=0.4\textwidth]{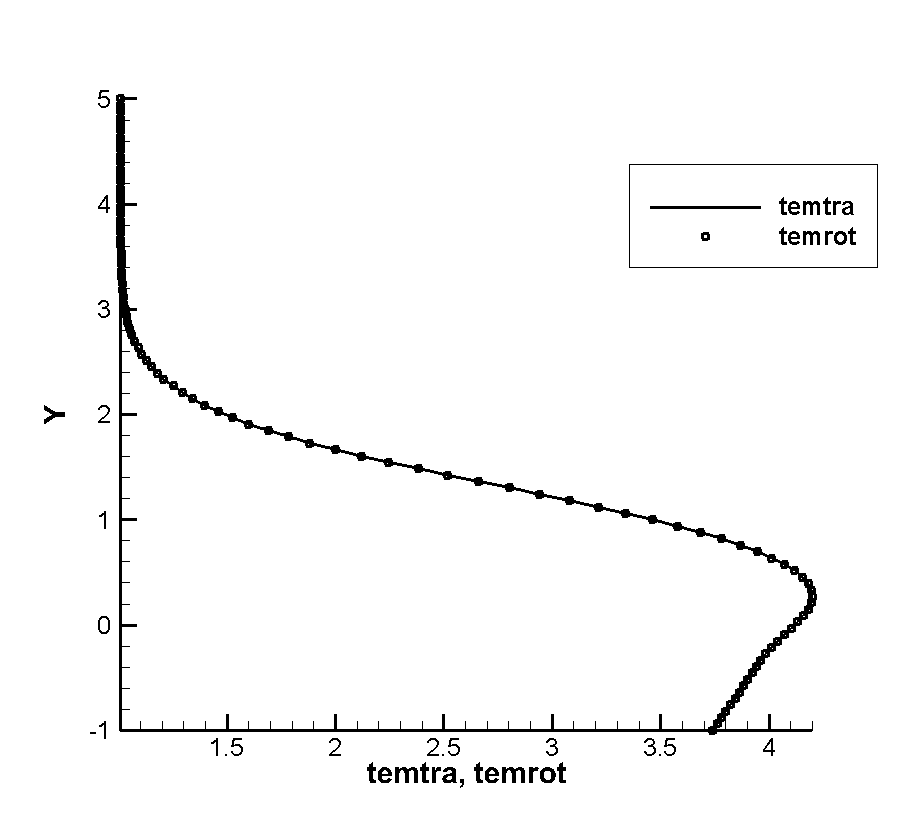}}
	\subfloat[X=308mm]{\includegraphics[width=0.4\textwidth]{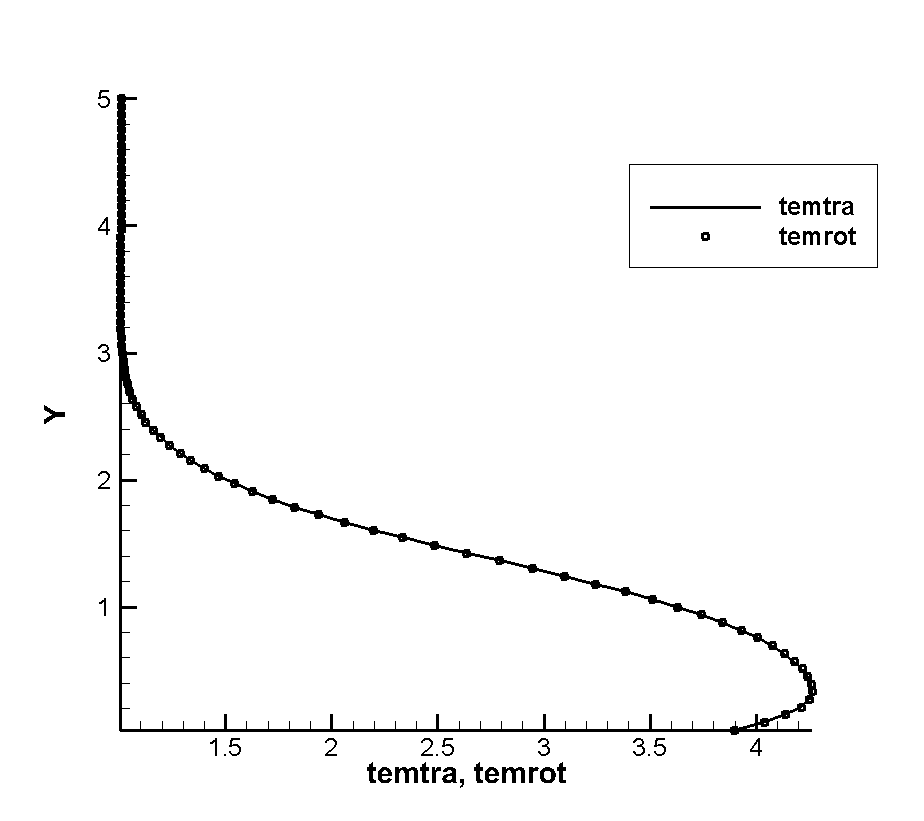}}
	\caption{Translational and rotational temperature profiles at wave peaks and troughs of corrugated wall W2}
	\label{3-20}
\end{figure}

\begin{figure}[H]
	\centering
	\includegraphics[width=0.5\textwidth]{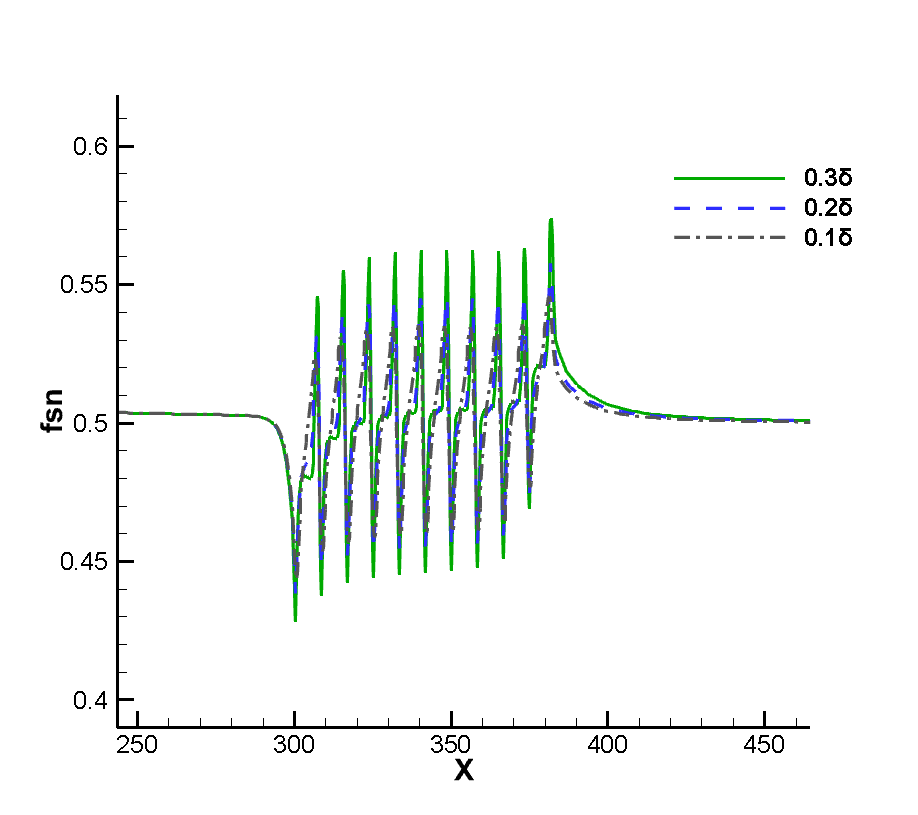}
	\caption{\label{3-21}Corrugation wall pressure}
\end{figure}

\begin{figure}[H]
	\centering
	\includegraphics[width=0.5\textwidth]{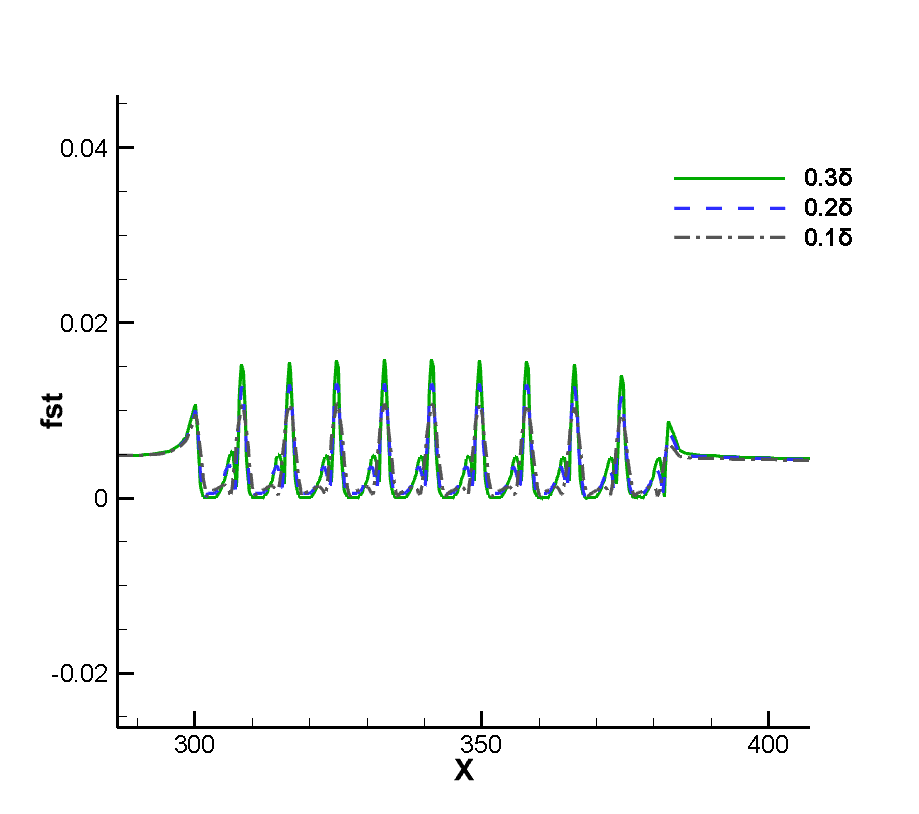}
	\caption{\label{3-22}Wall shear stress of corrugated wall}
\end{figure}

\begin{figure}[H]
	\centering
	\includegraphics[width=0.5\textwidth]{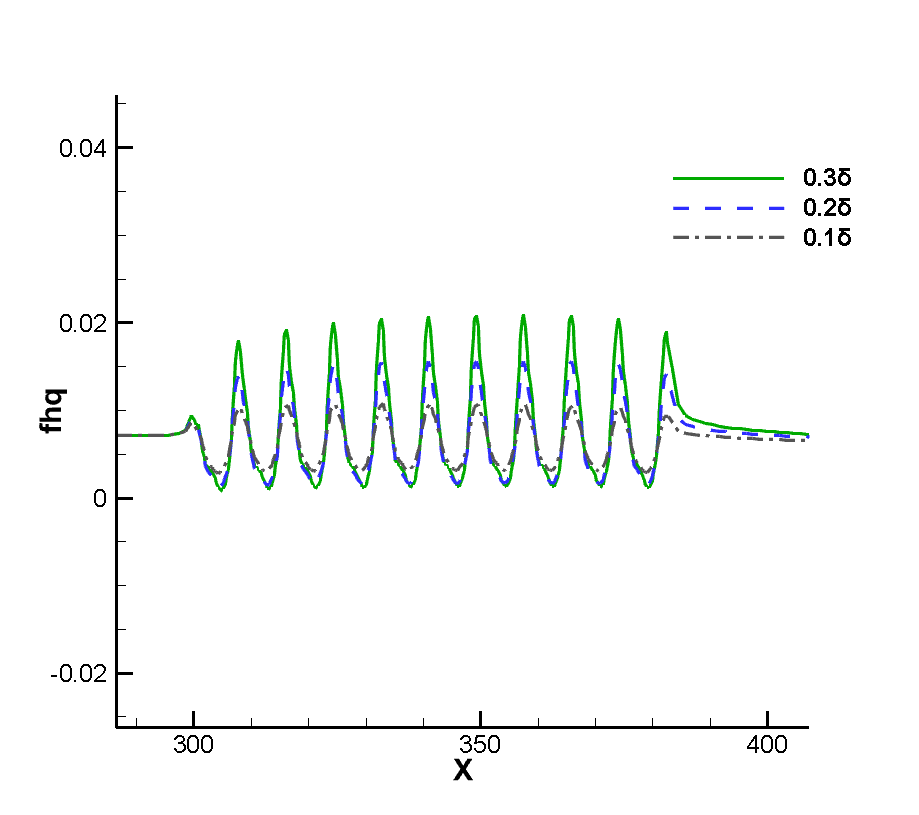}
	\caption{\label{3-23}Corrugation wall heat flow}
\end{figure}

\begin{figure}[H]
	\centering
	\subfloat[W1]{\includegraphics[width=0.3\textwidth]{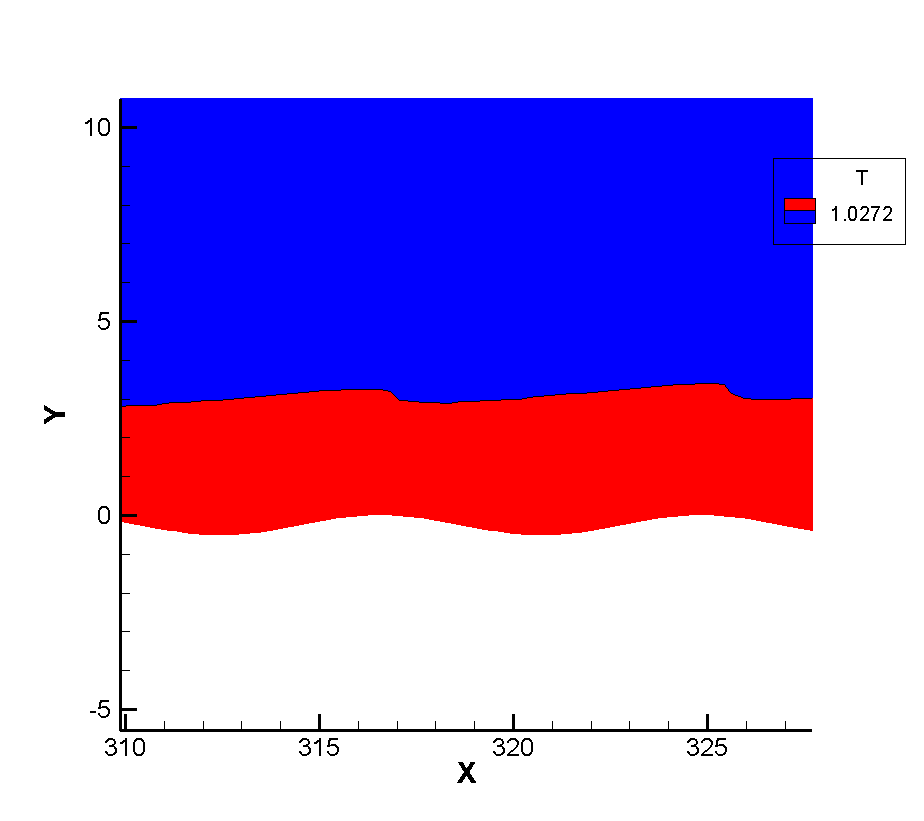}}
	\subfloat[W2]{\includegraphics[width=0.3\textwidth]{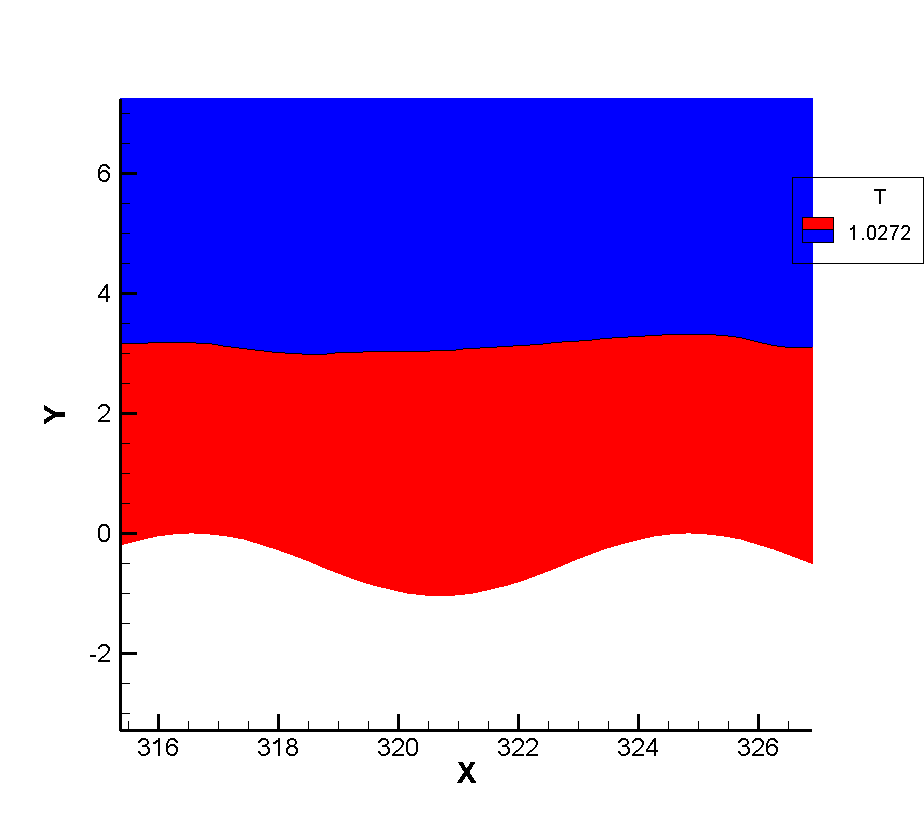}}
	\subfloat[W3]{\includegraphics[width=0.3\textwidth]{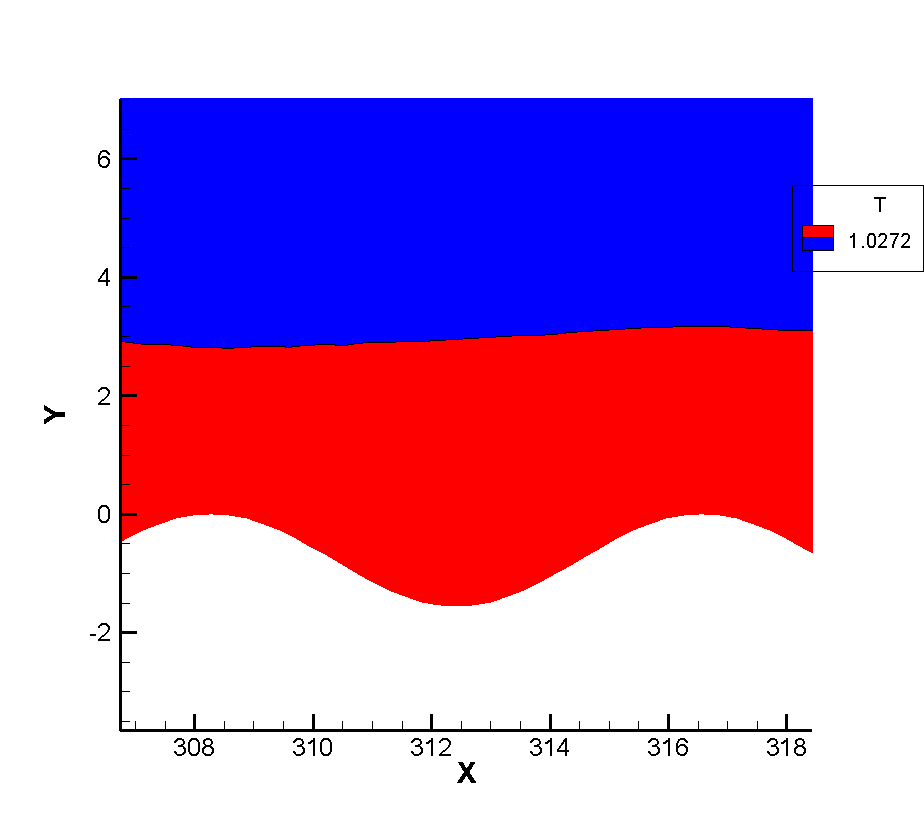}}
	
	\caption{Temperature boundary layer near the corrugated wall }
	\label{3-24}
\end{figure}

\begin{figure}[H]
	\centering
	\subfloat[$Re10^6$]{\includegraphics[width=0.3\textwidth]{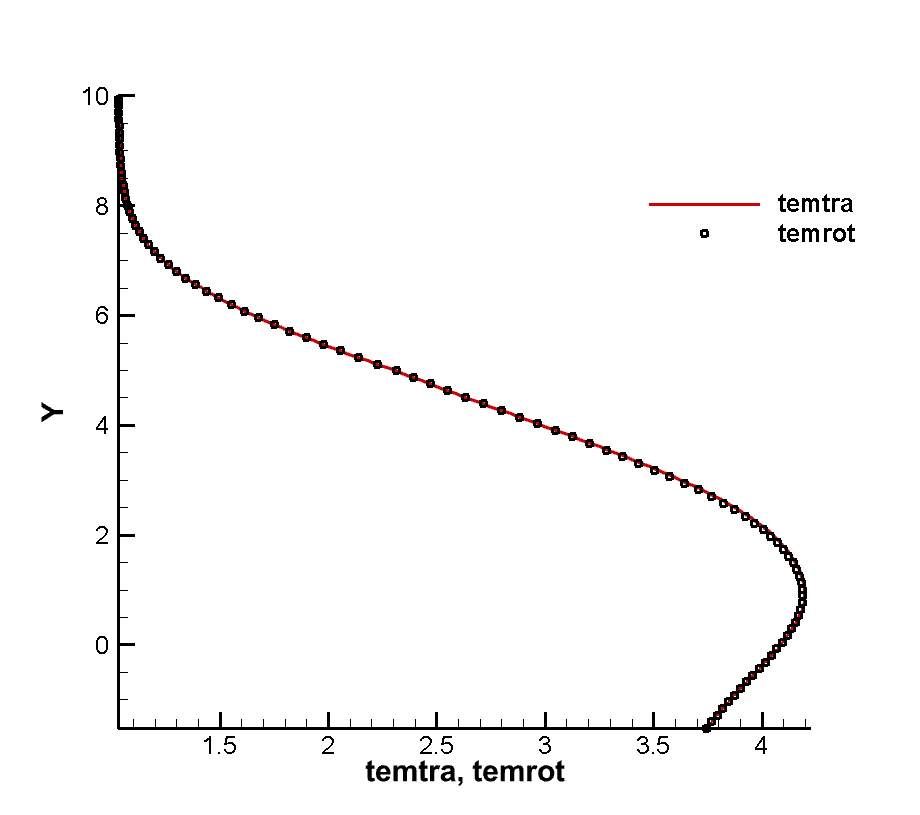}}
	\subfloat[$Re10^5$]{\includegraphics[width=0.3\textwidth]{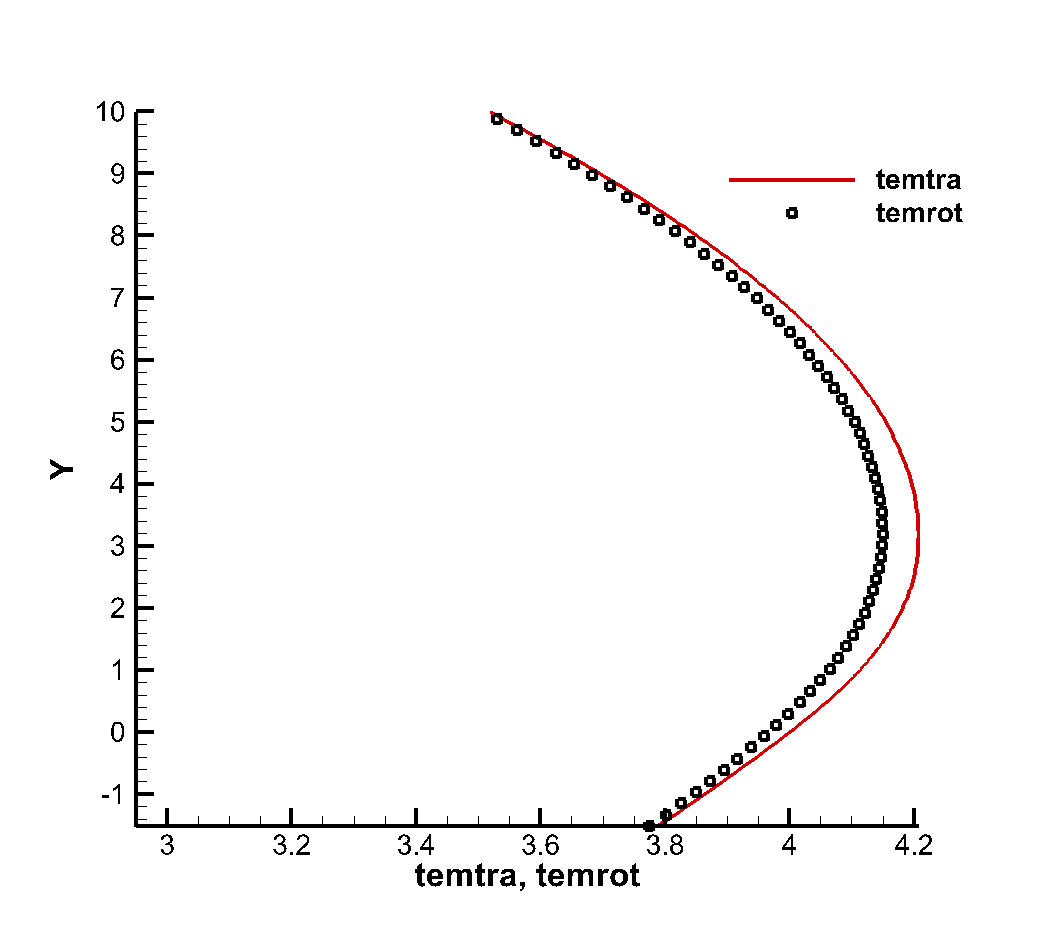}}
	\subfloat[$Re10^4$]{\includegraphics[width=0.3\textwidth]{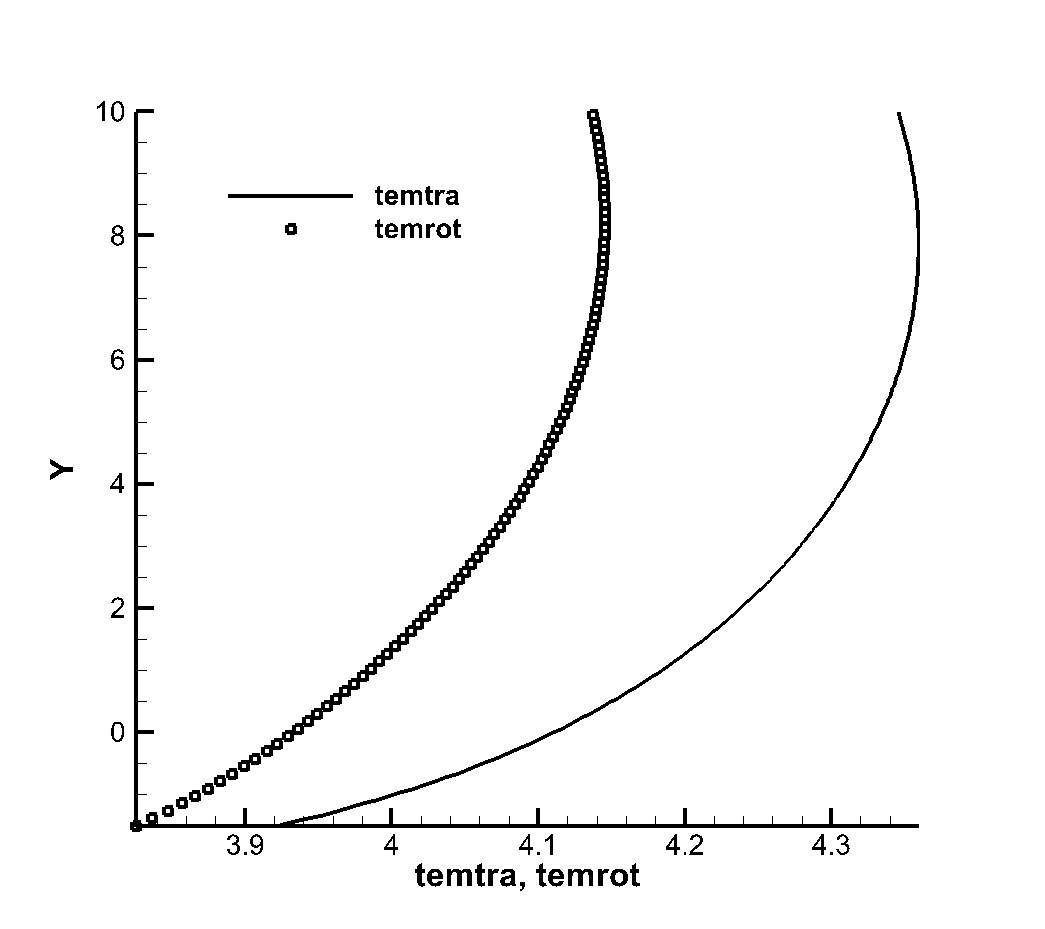}}
	
	\caption{Translational and rotational temperature profiles at the first trough }
	\label{3-25}
\end{figure}

\begin{figure}[H]
	\centering
	\subfloat[$Re10^7$]{\includegraphics[width=0.4\textwidth]{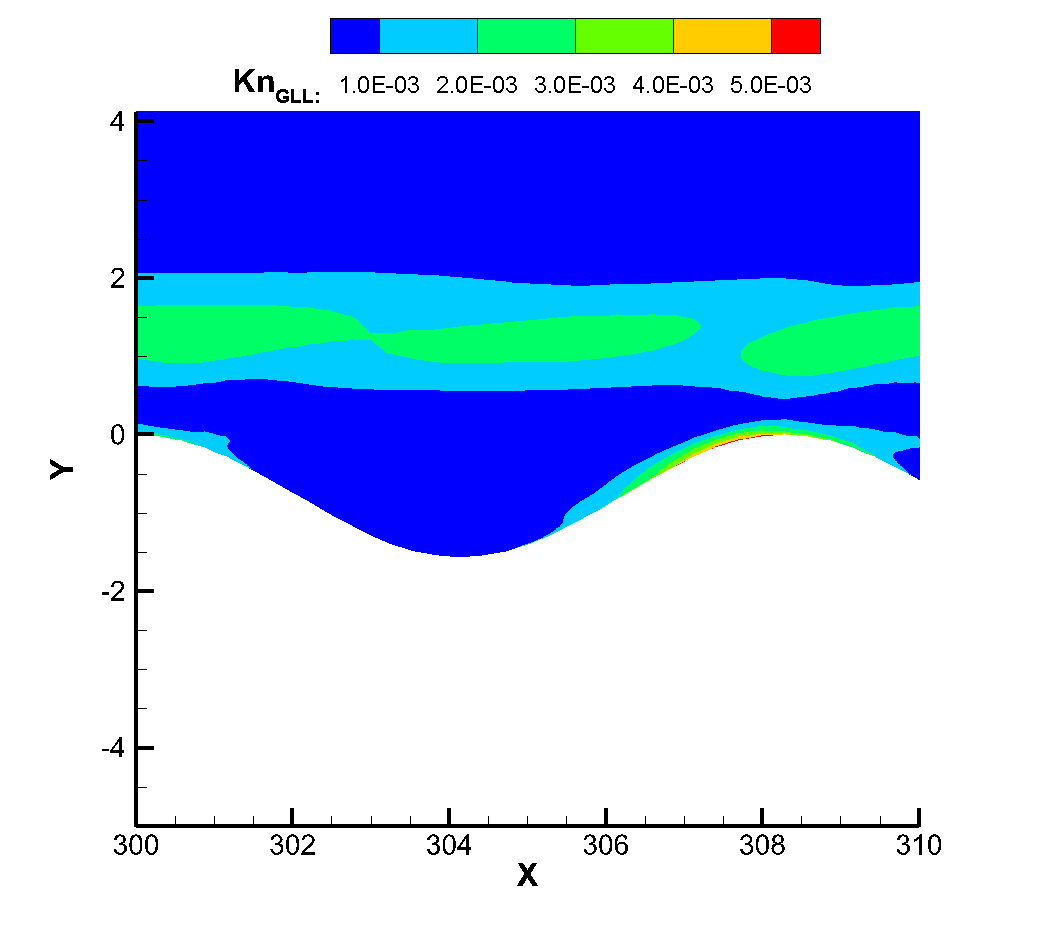}}
	\subfloat[$Re10^6$]{\includegraphics[width=0.4\textwidth]{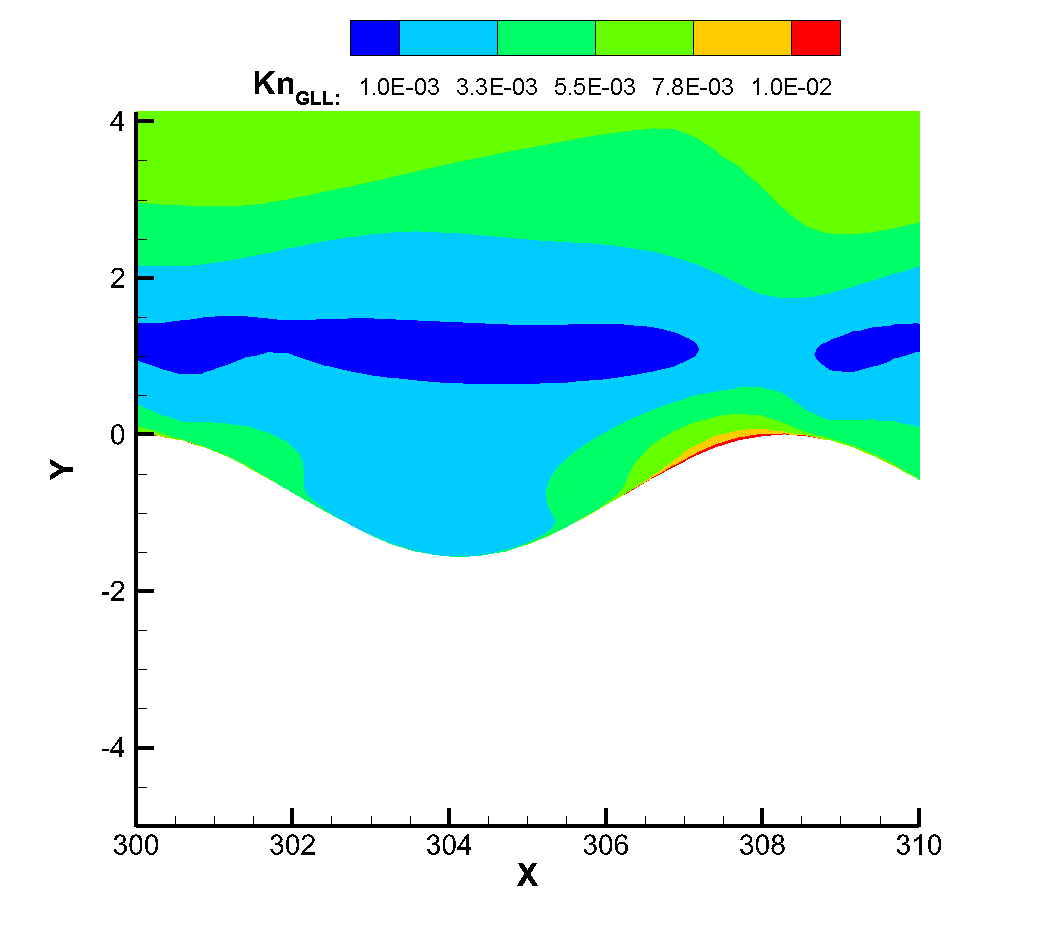}}
 \\
	\subfloat[$Re10^5$]{\includegraphics[width=0.4\textwidth]{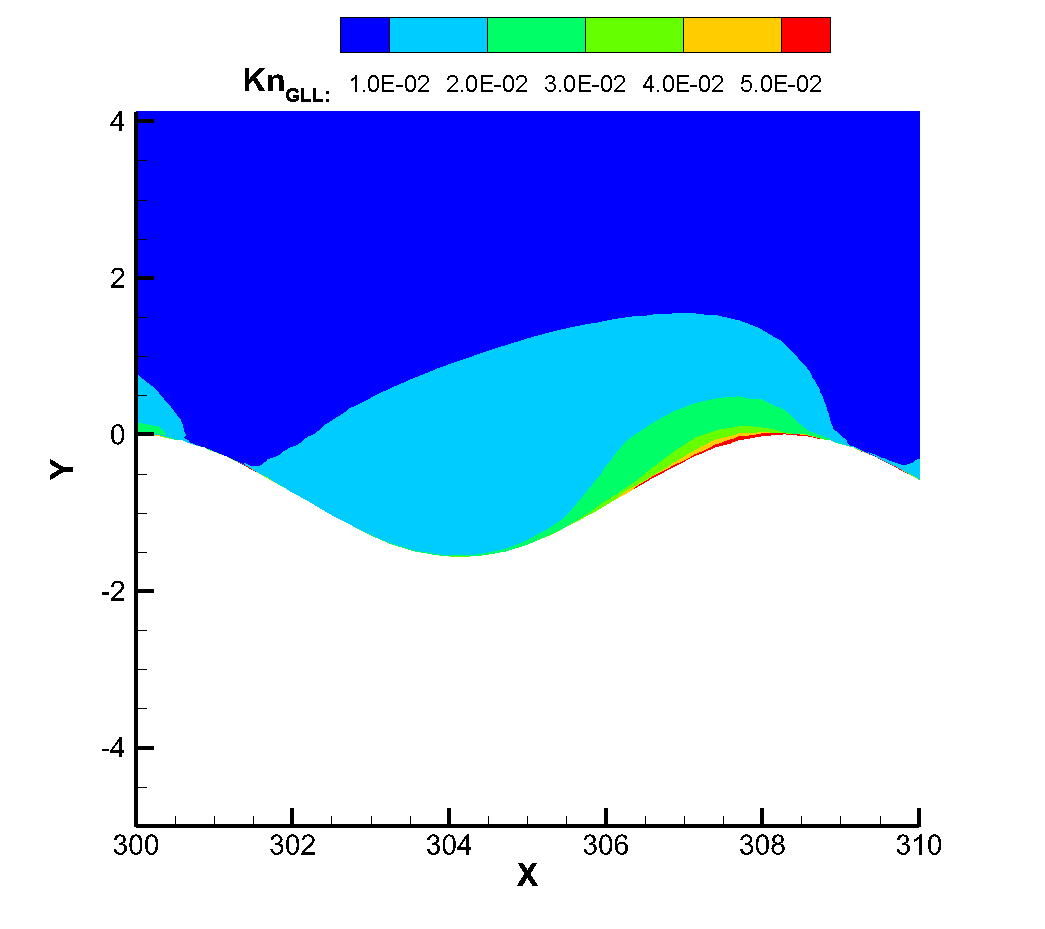}}
	\subfloat[$Re10^4$]{\includegraphics[width=0.4\textwidth]{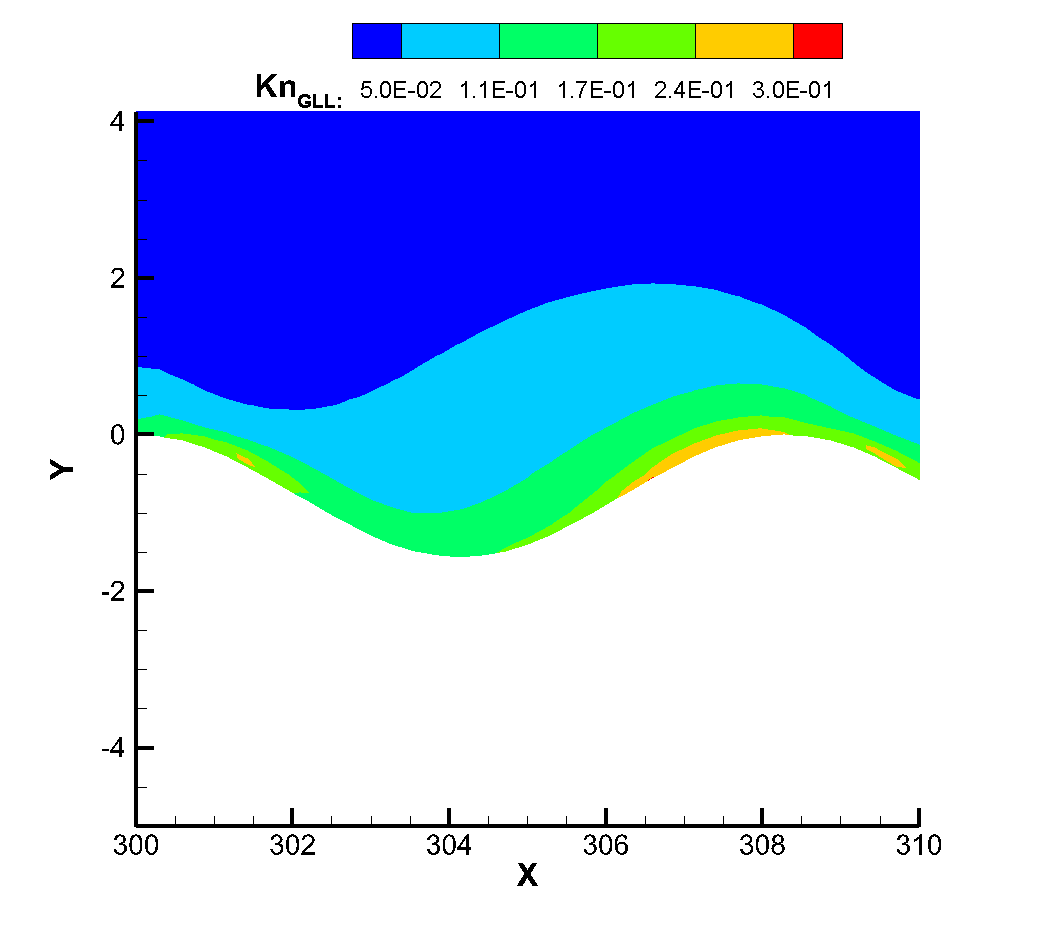}}
	\caption{Local Knudsen number near the corrugated wall}
	\label{3-26}
\end{figure}

\begin{figure}[H]
	\centering
	\subfloat[Velocity mesh]{\includegraphics[width=0.4\textwidth]{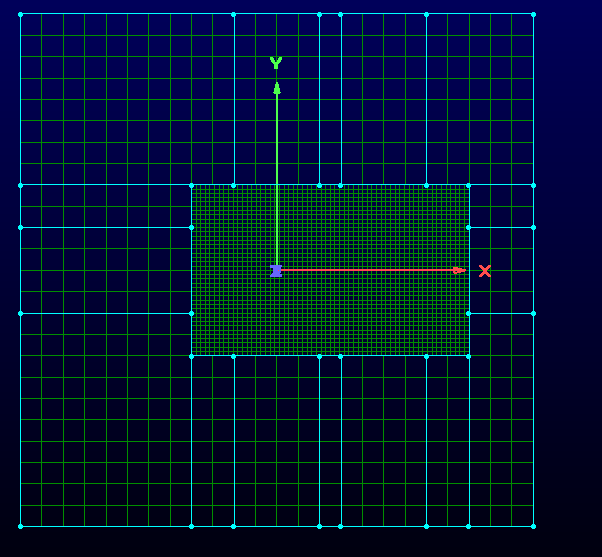}}
	\subfloat[Physical mesh]{\includegraphics[width=0.4\textwidth]{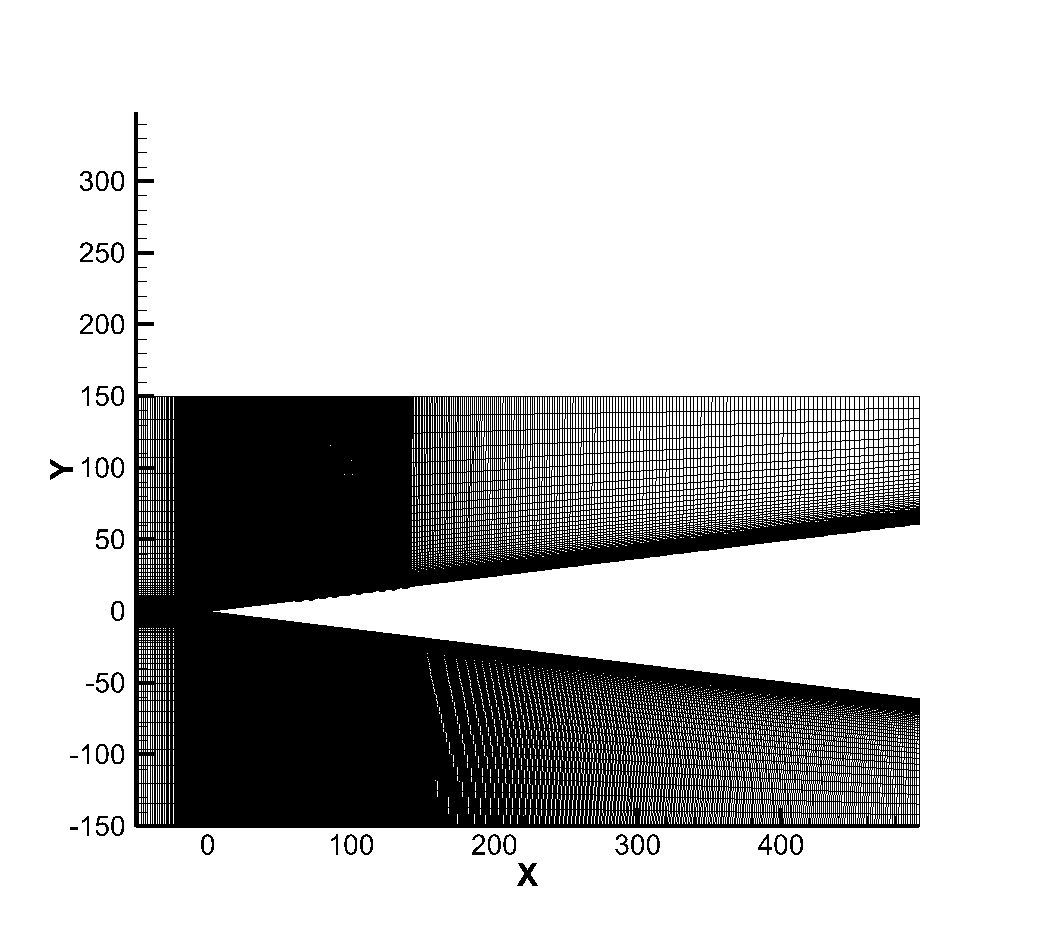}}

	\caption{The Reynolds number ${{10}^{7}}$ example computes the grid }
	\label{3-27}
\end{figure}

\begin{figure}[H]
	\centering
	\subfloat[$Re10^7$]{\includegraphics[width=0.4\textwidth]{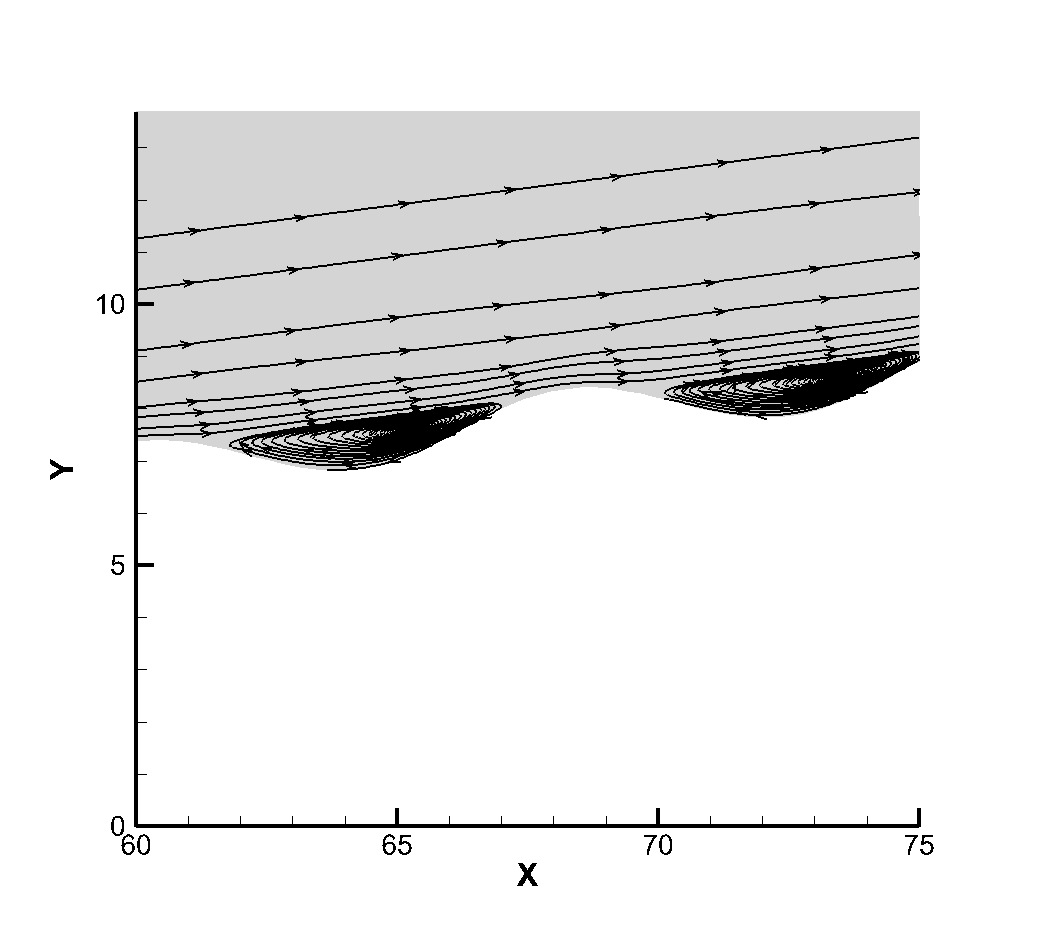}}
	\subfloat[$Re10^6$]{\includegraphics[width=0.4\textwidth]{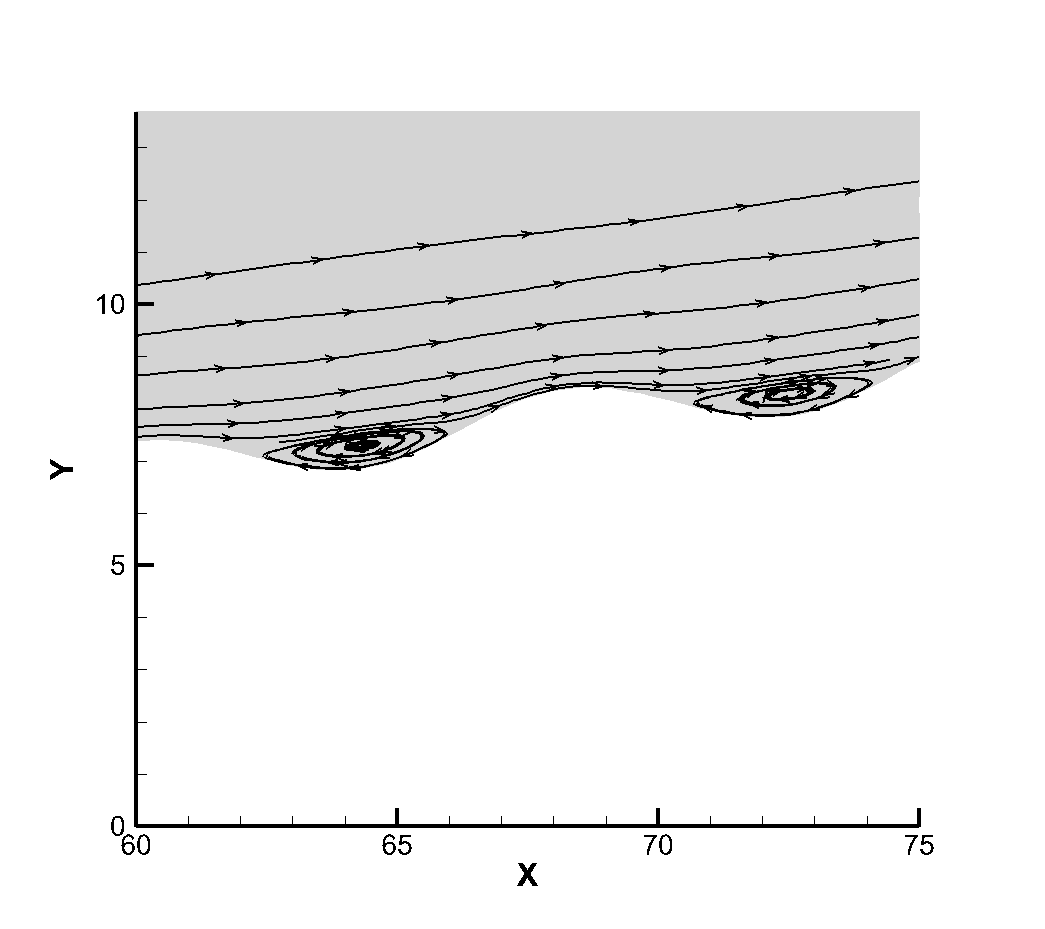}}
  \\
	\subfloat[$Re10^5$]{\includegraphics[width=0.4\textwidth]{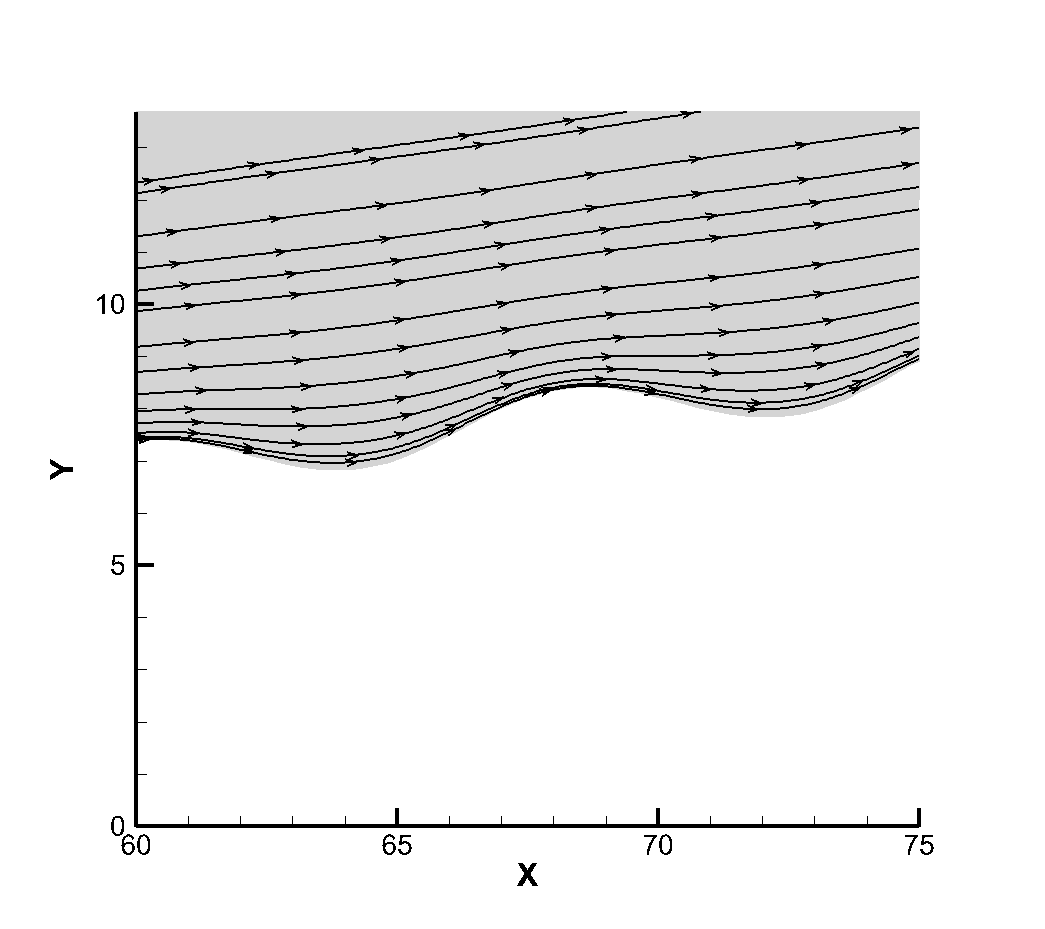}}
	\subfloat[$Re10^4$]{\includegraphics[width=0.4\textwidth]{3-28c}}
	\caption{Flow diagram near the corrugated wall }
	\label{3-28}
\end{figure}

\begin{figure}[H]
	\centering
        \subfloat[$Re10^7$]{\includegraphics[width=0.4\textwidth]{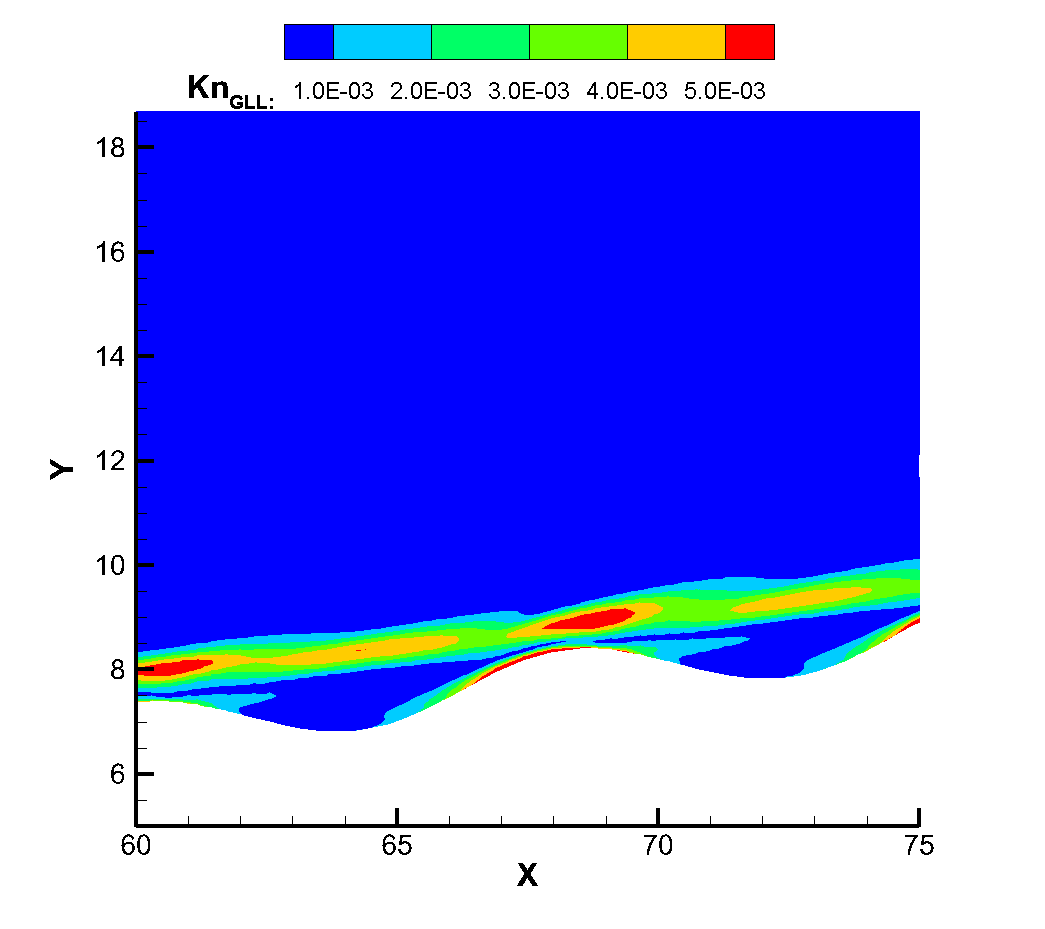}}
	\subfloat[$Re10^6$]{\includegraphics[width=0.4\textwidth]{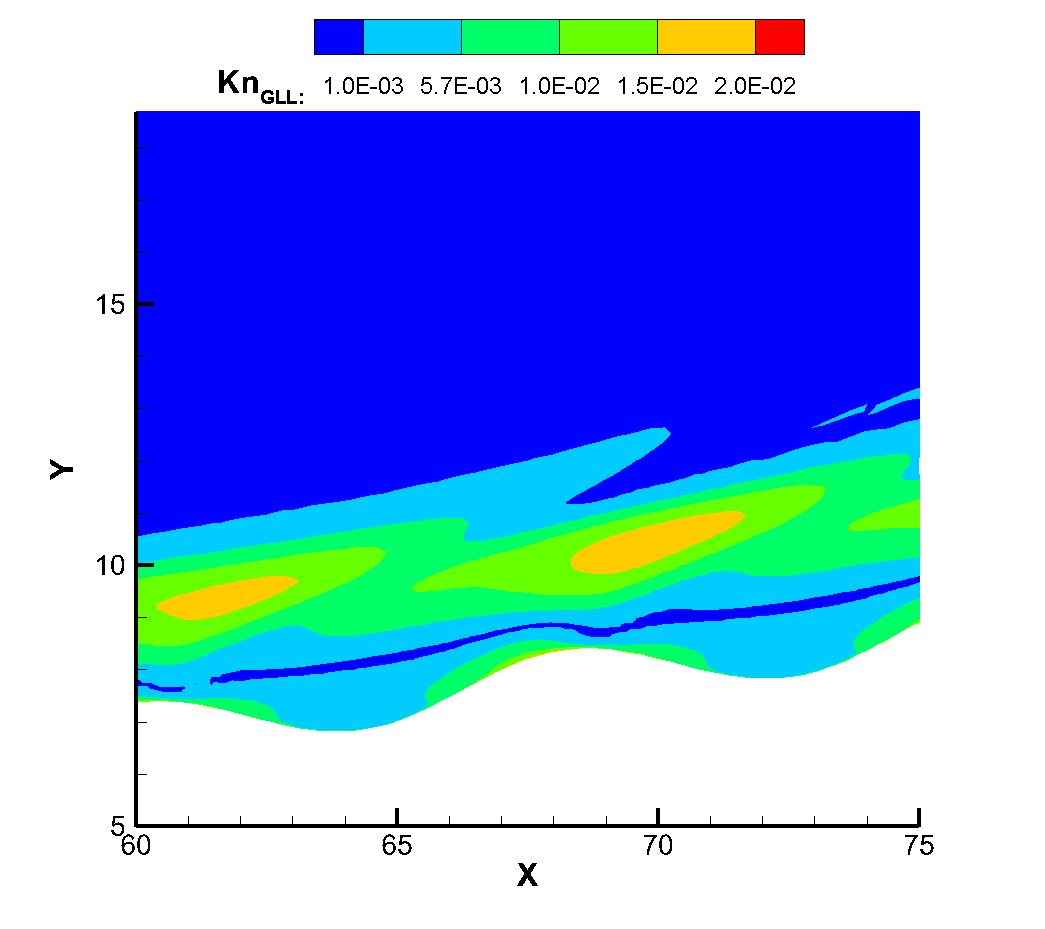}}
 \\
	\subfloat[$Re10^5$]{\includegraphics[width=0.4\textwidth]{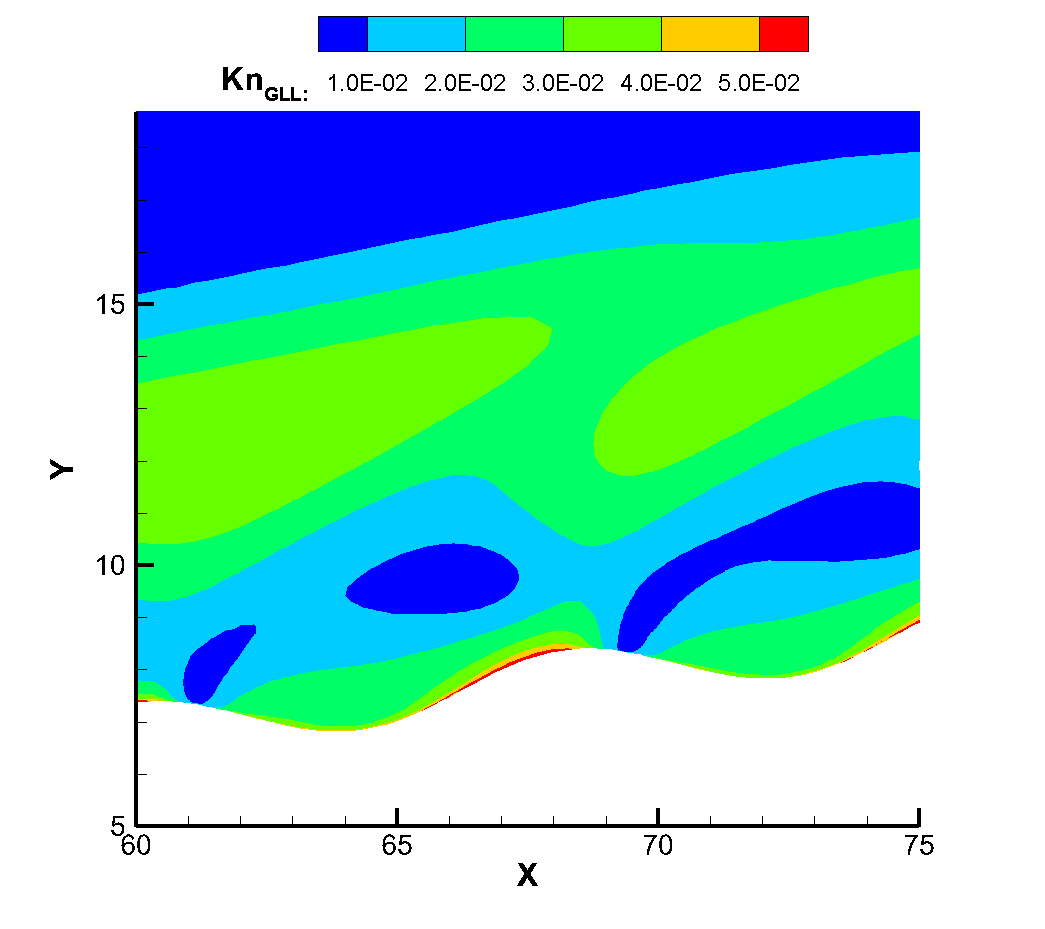}}
		\subfloat[$Re10^4$]{\includegraphics[width=0.4\textwidth]{3-29c}}
	\caption{Local Knudsen number near the corrugated wall}
	\label{3-29}
\end{figure}

\begin{figure}[H]
	\centering
	\subfloat[$Re10^7$]{\includegraphics[width=0.4\textwidth]{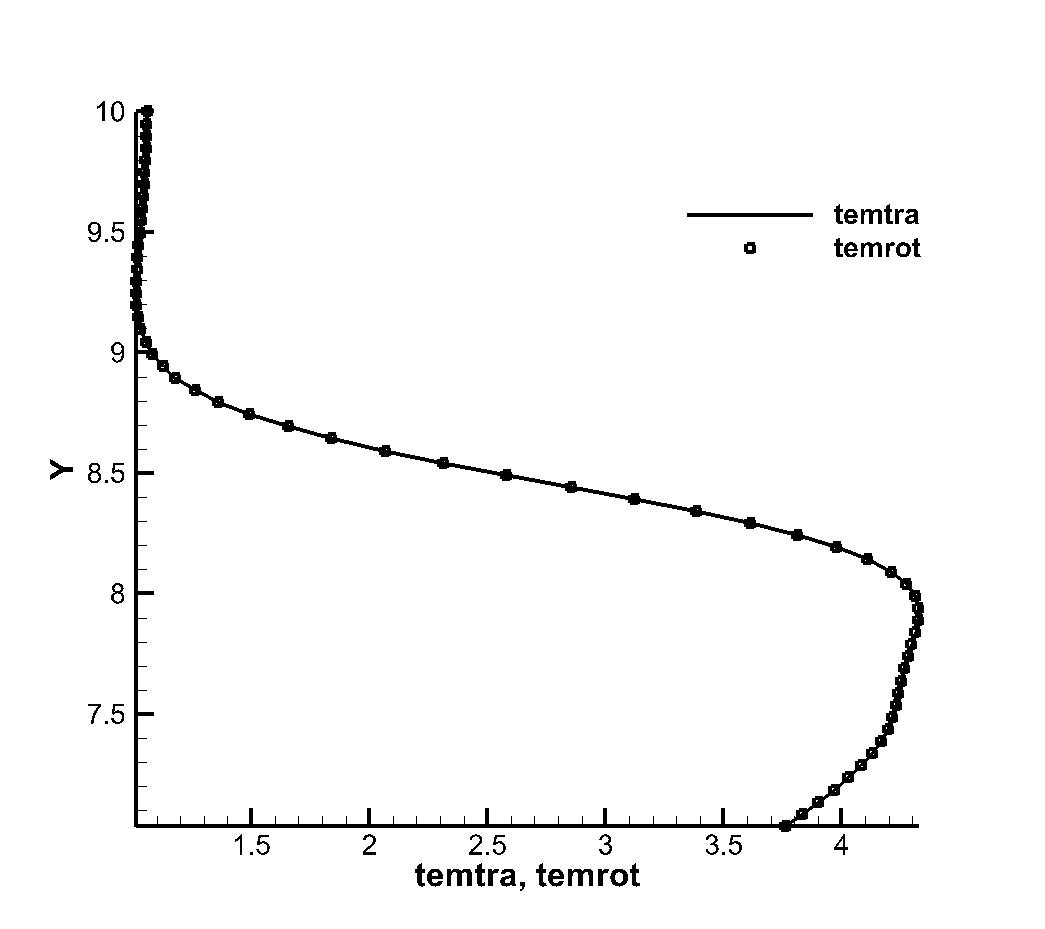}}
	\subfloat[$Re10^6$]{\includegraphics[width=0.4\textwidth]{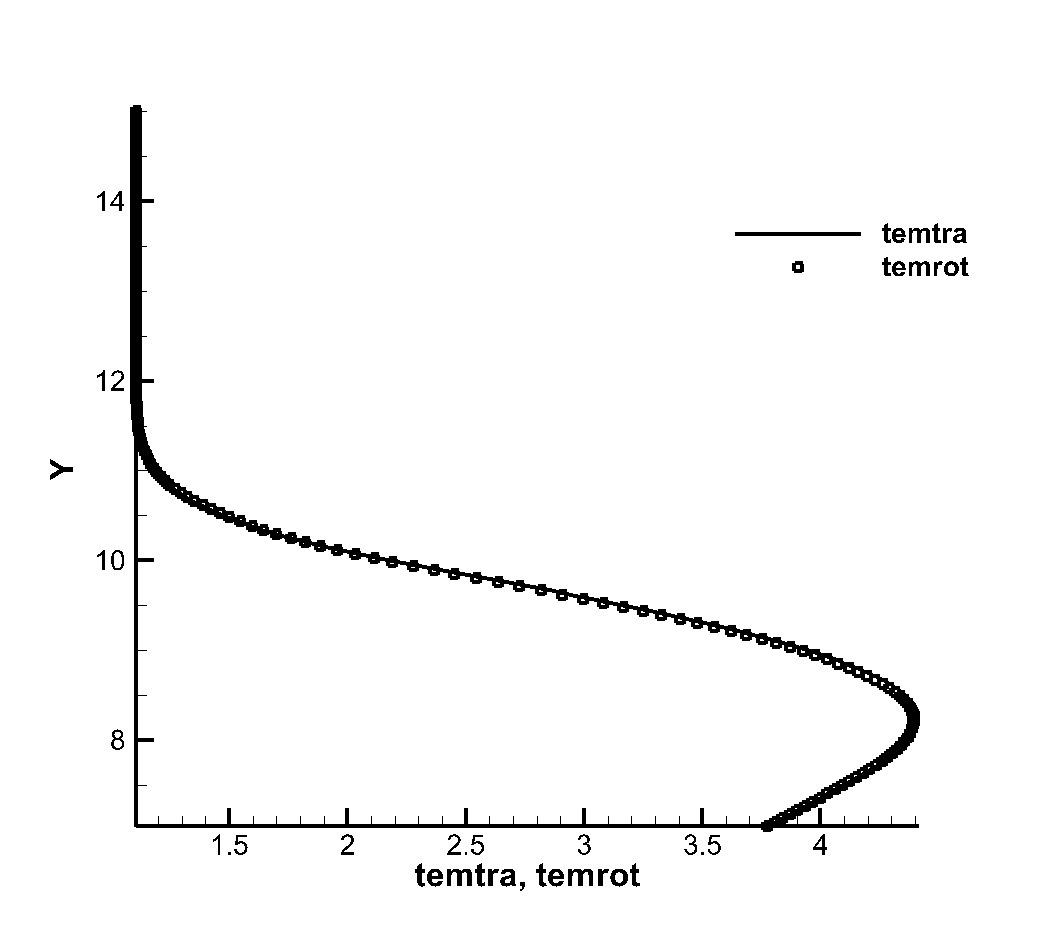}}
 \\
	\subfloat[$Re10^5$]{\includegraphics[width=0.4\textwidth]{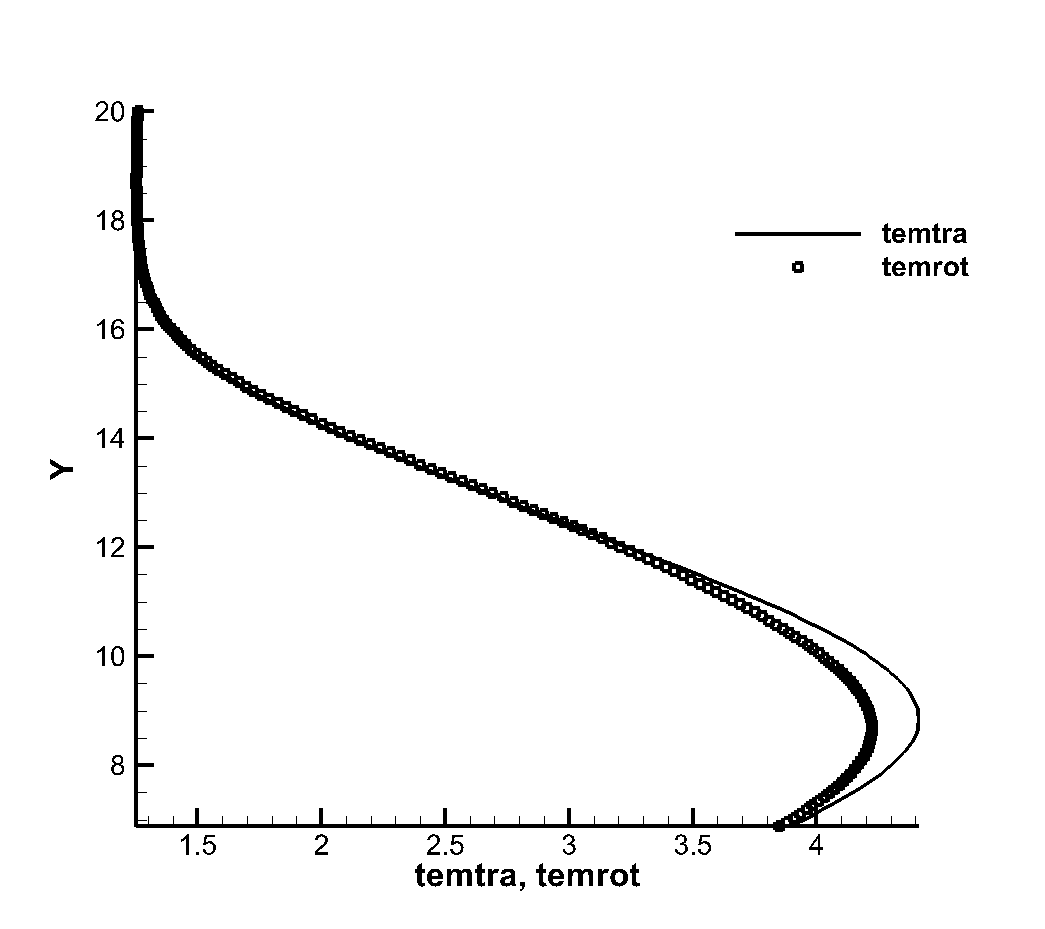}}
		\subfloat[$Re10^4$]{\includegraphics[width=0.4\textwidth]{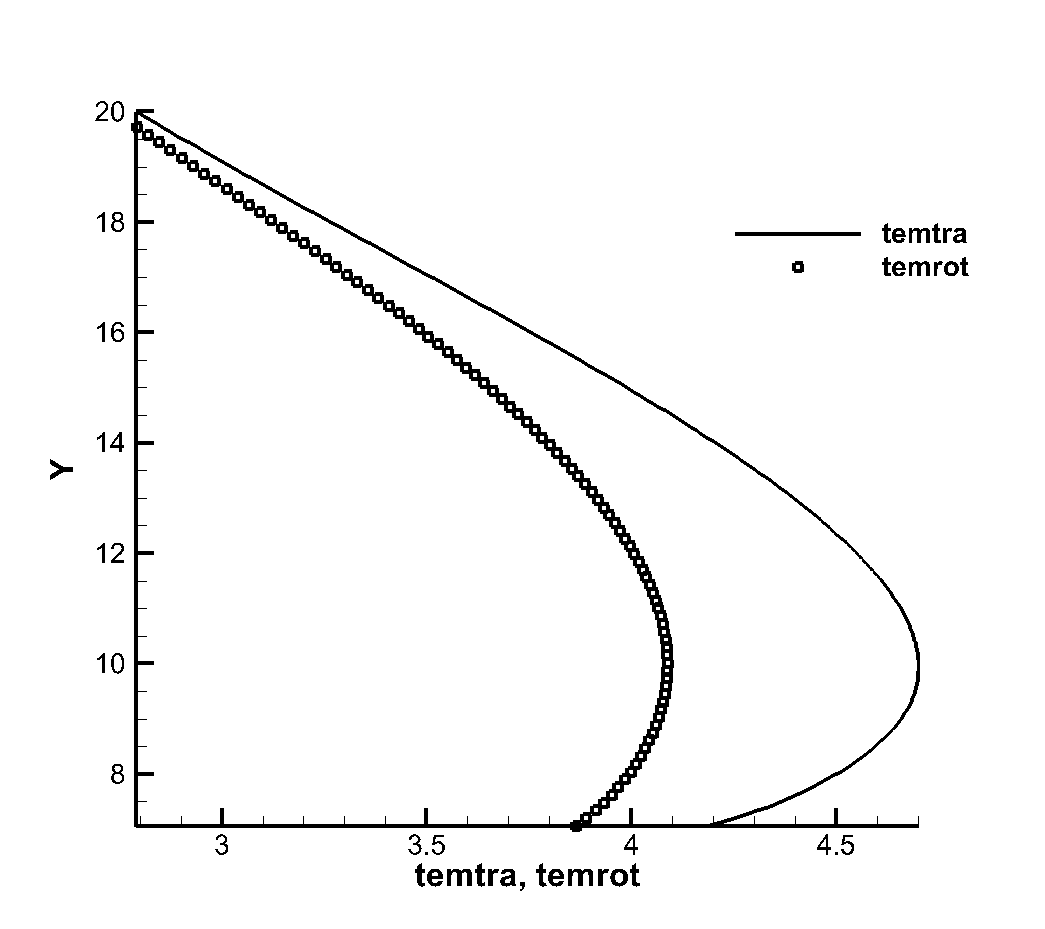}}
	\caption{Translational and rotational temperature profiles near corrugated walls}
	\label{3-30}
\end{figure}

	\bibliography{mybibfile}

\end{document}